\documentclass{article} 

\usepackage{jheppub} 

\usepackage{bbm} 
\usepackage{bm} 
\usepackage{hyperref}
\usepackage{amsmath,amssymb,amsthm} 
\usepackage{tensor} 
\usepackage{braket} 
\usepackage{parskip}
\usepackage[most]{tcolorbox}
\usepackage{ytableau}
    \ytableausetup{centertableaux}

\usepackage{tikz}
    \usetikzlibrary{
    tikzmark,
    positioning, 
    decorations.pathreplacing,
    calligraphy,
    matrix
    }

\usepackage[normalem]{ulem}    

\allowdisplaybreaks 

\hypersetup{
    linkcolor=[rgb]{0.91, 0.24, 0.43}, 
    citecolor=[rgb]{0.28, 0.24, 0.75}
    }

\newcommand{\dd}{\text{d}}
\newcommand{\pp}{\partial}
\newcommand{\DD}{\text{D}}
\newcommand{\ii}{\mathrm{i}}

\renewcommand{\[}{\left[}
\renewcommand{\]}{\right]}
\renewcommand{\(}{\left(}
\renewcommand{\)}{\right)}
\renewcommand{\>}{\right\rangle}
\renewcommand{\epsilon}{\varepsilon}
\newcommand{\<}{\left\langle}
\newcommand{\spar}{\underline{\star}}
\newcommand{\ol}{\overline}
\newcommand{\ul}{\underline}
\renewcommand{\ii}{\mathrm{i}}

\newtheorem{identity}{Identity}

\title{\centering Meet biform singletons in your area! \\
Topological actions for edge modes}

\author[a]{Calvin Y.-R. Chen,}
\author[b]{Euihun Joung,}
\author[c]{and Karapet Mkrtchyan}

\affiliation[a]{Max Planck--IAS--NTU Center for Particle Physics, Cosmology and Geometry, \\
Leung Center for Cosmology and Particle Astrophysics, and \\
Department of Physics \& Center for Theoretical Physics, National Taiwan University, Taipei 10617, Taiwan}
\affiliation[b]{Department of Physics, College of Science and \\
Research Institute for Basic Sciences,\\ 
Kyung Hee University, Seoul 02447, Republic of Korea}
\affiliation[c]{Abdus Salam Centre for Theoretical Physics, \\
Imperial College London, SW7 2AZ, United Kingdom}

\emailAdd{cyrchen@ntu.edu.tw}
\emailAdd{euihun.joung@khu.ac.kr}
\emailAdd{k.mkrtchyan@imperial.ac.uk}

\abstract{
We present topological field theories propagating biform singleton fields as edge modes.
In even spacetime dimensions $D=2p$, these are irreducible representations of the conformal algebra $\mathfrak{so}(2,2p)$ with two-column Young diagrams of height $p$. 
Their duality properties depend on the parity of $p$: They propagate chiral degrees of freedom for odd $p$, and exhibit electric--magnetic duality for even $p$.
We are able to construct the topological actions by realising the hidden conformal symmetry inherent to singletons geometrically in terms of an auxiliary bulk $\text{AdS}_{2p+1}$ geometry.
Using these as a starting point, we are able to apply a recently established covariant boundary reduction procedure to yield novel action formulations with manifest Lorentz covariance.
We also show how, breaking manifest Lorentz covariance, we recover known formulations of biform singletons.
}

\date{\today}

\begin{document}

\maketitle

\newpage 

\section{Introduction}

It is well known that particles and fields in $D$-dimensional Minkowski spacetime are in one-to-one correspondence with unitary irreducible representations (UIRs) of the Poincar\'e group $\text{ISO}(1,D-1)$.
In Wigner's classification, these representations fall into four classes: Massive, massless, tachyonic, and zero-momentum.
Massless representations are labelled by an irreducible representation (irrep) of the little group $\text{ISO}(D-2)$, and the usual massless helicity representations are those on which only the $\text{SO}(D-2)$ part acts non-trivially.
Among them sits, in even dimensions $D=2p$ ($p \in \mathbb{Z}_{>0}$), a distinguished class that also carries irreducible representations of the larger conformal group $\text{SO}(2,2p)$~\cite{Siegel:1988gd, Angelopoulos:1997ij, Laoues:1998ik, Angelopoulos:1999bz}.
These fields are known as \textbf{singletons}.

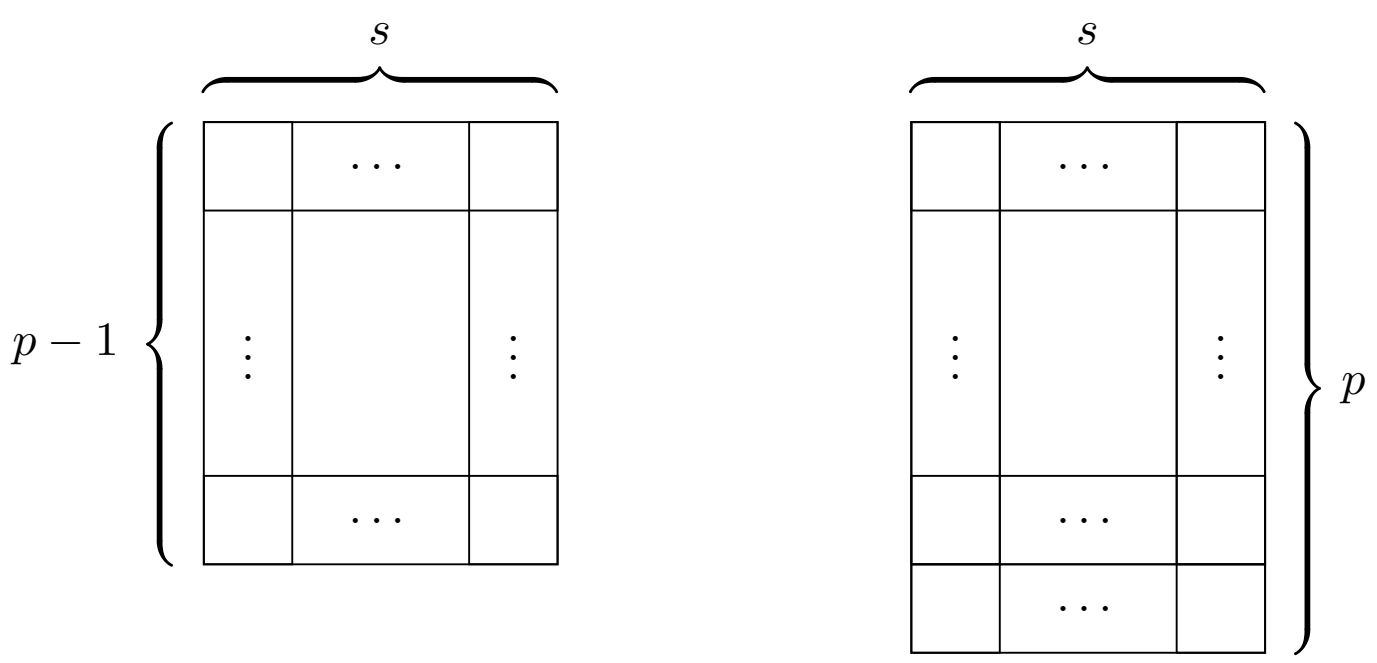
\begin{figure}
    \centering
    \scalebox{0.9}{\tikzset{
    cell/.style={
        anchor=south west,
        draw,
        minimum size=0.65cm,  
    },
    BC/.style = {decorate, 
        decoration={calligraphic brace, amplitude=5pt, raise=1mm},
        very thick, 
        pen colour={black}
    },                
}

\begin{tikzpicture}

    \node[cell, minimum width=4*0.65cm] at (0,1*0.65) {$\dots$};    
    \node[cell, minimum width=4*0.65cm] at (0,5*0.65) {$\dots$};    
    \node[cell, minimum height=5*0.65cm] at (0,1*0.65) {$\vdots$};
    \node[cell, minimum height=5*0.65cm] at (3*0.65,1*0.65) {$\vdots$};
    
    \node[cell, minimum width=4*0.65cm] at (8*0.65,0) {$\dots$};    
    \node[cell, minimum width=4*0.65cm] at (8*0.65,1*0.65) {$\dots$};    
    \node[cell, minimum width=4*0.65cm] at (8*0.65,5*0.65) {$\dots$};    
    \node[cell, minimum height=6*0.65cm] at (8*0.65,0*0.65) {};
    \node[cell, minimum height=6*0.65cm] at (11*0.65,0) {};
    \node[cell, minimum height=5*0.65cm] at (8*0.65,1*0.65) {$\vdots$};
    \node[cell, minimum height=5*0.65cm] at (11*0.65,1*0.65) {$\vdots$};

    \draw[BC] (-0.2*0.65,1*0.65) -- node[xshift=-2.5em] {$p-1$} (-0.2*0.65,6*0.65);
    \draw[BC] (0,6.2*0.65) -- node[yshift=1.5em] {$s$} (4*0.65,6.2*0.65);
    \draw[BC] (12.2*0.65,6*0.65) -- node[xshift=1.5em] {$p$} (12.2*0.65,0);
    \draw[BC] (8*0.65,6.2*0.65) -- node[yshift=1.5em] {$s$} (12*0.65,6.2*0.65);

\end{tikzpicture}}
    \caption{The fields considered in this manuscript have window-type Young diagrams.
    In $D=2p$ dimensions, (anti-)self-duality and twisted self-duality relations are possible for field strengths with Young symmetry $[p^{s}]$, shown on the right.
    The diagram on the left shows the corresponding fundamental gauge potential field.}
    \label{fig: window}
\end{figure}

As representations of the little group $\text{SO}(2p-2)$, singletons are labelled by rectangular Young diagrams of height $p-1$ and arbitrary width $s$ (the spin)~\cite{Siegel:1988gd, Bandos:2005mb, Bekaert:2009fg, Barnich:2015tma}.\footnote{Half-integer $s$ is also allowed ~\cite{Ferrara:2000nu}. 
These fermionic singletons are realised by Weyl spinor--tensors, whose potentials have tensor part $[(p-1)^{s-1/2}]$~\cite{Bandos:2005mb}.
Examples are the gravitino ($s=\tfrac{3}{2}$) in $D=4$ and the exotic gravitino in $D=6$~\cite{Hull:2000zn, Henneaux:2017xsb}. 
In this manuscript, we restrict to bosonic singletons.}
At the level of \textit{fields}, they are realised by $\text{SO}(1,D-1)$ tensor potentials with the same Young diagrams $[(p-1)^{s}]$.\footnote{Here, we use $[\cdot]$ to denote the column heights of a Young diagram.} 
Their field strengths carry rectangular Young diagrams with one additional row, giving height $p$ and width $s$.
Both are shown in figure~\ref{fig: window}.
Since, up to dualisation, an $\text{SO}(N)$ Young diagram has at most $N/2$ rows (for even $N$), the Young diagrams of both the little-group representation and the field strength are extremal: They saturate this bound.
	
The dynamics of singletons are encoded by \textbf{(twisted) self-duality} relations, which connect their field strength tensors to their Hodge duals.
The Hodge dual acts on a single column of indices, replacing its height $p$ by $D-p$.
In $D=2p$ dimensions, the field strength and its dual therefore have the same rectangular Young diagram.
Precisely which kind of first-order relation applies depends on the spacetime signature and dimension.
In Lorentzian signature, the Hodge dual acting on the field strength satisfies
\begin{equation}
    \star^{2} = (-)^{p+1}\,,
\end{equation}
so the structure of the equations depends on the value of $p\text{ mod }2$:
\begin{itemize}
    \item Odd $p=2k+1$: In this case $\star^{2} =1$, and the Hodge dual $\star$ defines an \textit{involution}.
    The space of field strengths splits into two chiral halves,
    \begin{equation}
        \mathcal{V} = \mathcal{V}_{-} \oplus \mathcal{V}_{+},
    \end{equation}
    with $\star$ acting on $\mathcal{V}_{\pm}$ as $\pm 1$ respectively.
    In $D=4k+2$ dimensions, a field strength without further constraints therefore contains two decoupled \textit{chiral} sectors.
    A singleton keeps only one of them, i.e. its field strength $F$ obeys a genuine (anti-)self-duality condition,    
    \begin{equation}
        \star F = \pm F.
    \end{equation}
    The best-known examples are the chiral scalar and  its higher-order generalisations in $D=2$ ($p=1$)~\cite{Floreanini:1987as, Chen:2025xlo}, which are closely related to discrete series scalar fields in (A)dS$_{2}$~\cite{Sengor:2022lyv, Sengor:2022kji, Anninos:2023lin, Farnsworth:2024yeh, Chen:2026kjo, Lindwasser:2026exa}. 
    Other well-known examples are the chiral forms in $D=6$ ($p=3$) \cite{Hull:2000rr, Hull:2000zn, Henneaux:2017xsb, Basile:2024raj} and $D=10$ ($p=5$) \cite{Schwarz:1983wa, Schwarz:1983qr, Howe:1983sra,  Henneaux:1988gg, Schwarz:1993vs, Pasti:1995tn, Pasti:1996vs, Sen:2015nph}.

    \item Even $p = 2k$: In this case $\star^{2} = -1$, and the Hodge dual $\star$ defines a \textit{complex structure}.
    The space of field strengths no longer splits, but its complexification does,
    \begin{equation}
        \mathcal{V} \otimes \mathbb{C} = \mathcal{V}^{\mathbb{C}}_{-} \oplus \mathcal{V}^{\mathbb{C}}_{+},
    \end{equation}
    with $\star$ acting on  $\mathcal{V}_{\pm}^{\mathbb{C}}$ as $\pm \ii$ respectively.
    A singleton again keeps only one of the two halves.
    In terms of real fields, this is described in terms of two field strengths related by a \textit{twisted} self-duality relation,
    \begin{equation}
        \star F^{(1)} = \pm F^{(2)},
    \end{equation}
    so that introducing the second field strength does not double the physical degrees of freedom.
    The best-known examples are massless spin-$s$ gauge fields in $D=4$ ($p=2$).
    Duality-symmetric actions for them were constructed in~\cite{Henneaux:1988gg, Schwarz:1993vs, Deser:1997se, Henneaux:2020nxi} for $s=1$, in~\cite{Henneaux:2004jw, Barnich:2008ts, Bunster:2011qp, Bunster:2013oaa, Henneaux:2016opm, Lekeu:2018kul, Henneaux:2019zod} for $s=2$, and \cite{Henneaux:2016zlu} for higher-spin.
\end{itemize}
For fixed $s=1$, the singleton family also recovers the full family of (anti-)chiral forms and their duality-symmetric counterparts~\cite{Siegel:1983es, Henneaux:1988gg, Tseytlin:1990va, Schwarz:1993vs}.
These illustrate a challenge common to the entire family: Even though these theories are well understood, constructing actions that make all their symmetries manifest requires novel formulations that go beyond the orthodox framework of gauge field and gravitational theories.\footnote{For massless gauge fields in four dimensions, for instance, electric--magnetic duality becomes manifest when the field equations are written as twisted self-duality relations.
Making it manifest at the level of the action, in turn, typically obscures Lorentz covariance.}
Such formulations are worth pursuing, since they provide a natural starting point for interactions or extensions that preserve these symmetries.
For chiral forms, several approaches now provide actions in which all these symmetries are manifest~\cite{Pasti:1995ii, Pasti:1995tn, Pasti:1995us, Pasti:1996vs, Maznytsia:1998xw, Miao:2000fn, Ivanov:2014nya, Sen:2015nph, Sen:2019qit, Mkrtchyan:2019opf, Avetisyan:2021heg, Evnin:2022kqn, Arvanitakis:2022bnr, Evnin:2023ypu, Hull:2023dgp,Ferko:2024zth}.
More recently,~\cite{Arvanitakis:2022bnr, Evnin:2023ypu} suggested that covariant boundary reductions of topological field theories (when those exist) give a systematic way to derive such actions.

A complementary perspective motivates this approach: The conformal group $\text{SO}(2,D)$ is also the isometry group of $(D+1)$-dimensional anti-de Sitter spacetime, AdS$_{D+1}$~\cite{Maldacena:1997re, Witten:1998qj}.\footnote{
Historically, singletons were first discovered precisely in this context.
Dirac discovered two `remarkable' representations of the AdS$_4$ isometry group $\text{SO}(2,3)$ that are naturally described on the three-dimensional boundary~\cite{Dirac:1963ta}.
Flato and Fronsdal later named them the \textit{Rac} and the \textit{Di}, and showed that their tensor products decompose into the tower of massless fields in AdS$_4$~\cite{Flato:1978qz}.
Singletons of higher-dimensional conformal algebras, including those of arbitrary spin, were subsequently classified in~\cite{Siegel:1988gd, Angelopoulos:1997ij, Laoues:1998ik, Angelopoulos:1999bz}.
The relation between masslessness in AdS$_D$ and in flat space was clarified in~\cite{Brink:2000ag}.
For higher-spin singletons  of $\mathfrak{so}(2,4)$, the analogous tensor product decomposition follows from the character formulae of~\cite{Dolan:2005wy}.
A pedagogical review of singletons and their maximal symmetry algebras is given in~\cite{Bekaert:2011js}.}
As in flat space, particles and fields in AdS$_{D+1}$ are in one-to-one correspondence with UIRs of $\text{SO}(2,D)$.
The singletons appear here as the extremal case once more: Their $\text{SO}(D)$ diagrams are rectangular with height $p$, which saturates the corresponding $\text{SO}(D)$ bound~\cite{Angelopoulos:1999bz, Bekaert:2011js}.\footnote{Such a representation is labelled by its ground-state energy $E_0=\Delta$ and the $\text{SO}(D)$ representation carried by the ground state.
Let the $\text{SO}(D)$ Young diagram be $[h_{1}^{s_{1}}, h_{2}^{s_{2}},\ldots,h_{k}^{s_{k}}]$, made of $k$ rectangular blocks of $s_i\times h_i$ boxes.
Unitarity then requires $E_0\ge s_1-h_1+D-1$~\cite{Metsaev:1995re}.
Massless fields sit at this unitarity bound, while massive fields lie strictly inside the unitary region~\cite{Metsaev:1995re, Brink:2000ag}.
For singletons, with $h_1=p$, the lowest energy $E_0=s+p-1$~\cite{Angelopoulos:1999bz, Bekaert:2011js} coincides with the value of this bound.}
However, from the perspective of AdS$_{D+1}$, singletons are topological: They carry no local degrees of freedom in the bulk and live purely on the conformal boundary.
This can be seen by restricting them to the isometry group of the $D$-dimensional boundary.
The height-$p$ diagram of $\text{SO}(D)$ branches to only two copies of the rectangular diagram of height $p-1$ in $\text{SO}(D-1)$.
Each copy corresponds to one of the (anti-)self-dual halves of the $\text{SO}(D)$ tensor: For odd $p$, each half by itself defines a real (chiral) singleton, whereas for even $p$ the two halves are complex conjugates of each other and a real field carries both.

This perspective suggests a natural approach: Constructing topological theories in one dimension higher whose edge modes are singletons.
This has already been established for individual members of the singleton family, such as the chiral scalar in two dimensions~\cite{Witten:1988hf} and, more generally, $p$-forms~\cite{Fuentealba:2023usq}.
Our goal is to extend this to the entire singleton family.
This is particularly challenging for spins $s>1$, because the relevant bulk fields are differential forms valued in non-trivial representations of the spacetime symmetries.
We initiated this programme in~\cite{Chen:2025xlo} with the higher-order chiral scalars in two dimensions\ ($p=1$ and $s$ arbitrary).
In~\cite{Chen:2026chi}, we subsequently extended this to linearised gravity in four dimensions ($s=2$ and $p=2$), presenting a description in terms of edge modes of a novel topological action in AdS$_{5}$.
A further covariant boundary reduction of this action then gave an action for $4d$ linearised gravity that is both manifestly Lorentz covariant and duality symmetric, resolving the obstruction previously identified in \cite{Henneaux:2004jw}.

\hspace{2em} In this manuscript, we take the programme further and generalise our construction to singletons with $s=2$ and \textit{arbitrary $p$}.
We refer to these as \textbf{biform singletons}.
Following~\cite{Chen:2025xlo, Chen:2026chi}, our central result is a family of topological field theories in AdS$_{2p+1}$ that propagate biform singletons as edge modes.
These bulk topological actions serve as parent actions, from which various boundary formulations can be derived.

At a technical level, these bulk theories rely on the frame-like formulation of gauge fields in AdS.
Two of its features are central for us: The realisation of mixed-symmetry fields by higher-degree forms and the structure of the action principle.
In the frame-like formulation, a massless field is described not by a single tensor field but by differential forms valued in finite-dimensional irreducible modules of the AdS algebra  $\mathfrak{so}(2,D)$.
For totally symmetric fields, these are one-forms: The generalised vielbein and connections combine into a single one-form valued in an irreducible $\mathfrak{so}(2,D)$-module, generalising the way the vielbein and spin connection of gravity form an AdS connection~\cite{Vasiliev:1980as, Lopatin:1987hz}.
For mixed-symmetry fields, this is no longer the case.
As shown in~\cite{Alkalaev:2003qv, Alkalaev:2005kw, Alkalaev:2006rw}, a mixed-symmetry field in AdS is described by a $q$-form valued in an $\mathfrak{so}(2,D)$-module $\mathcal{V}$ determined by its Young diagram,\footnote{The linearised curvatures and gauge transformations are generated by the covariant derivative $\DD$ of the AdS background, which squares to zero; the gauge symmetries are reducible, with $(q-1)$-form gauge parameters.
This higher-form realisation makes the Brink--Metsaev--Vasiliev phenomenon manifest: In AdS, only the gauge symmetry associated with one block survives, and the field content reorganises accordingly~\cite{Brink:2000ag, Boulanger:2008up, Boulanger:2008kw}.
Unfolded and BRST formulations of these fields are given in~\cite{Skvortsov:2009nv, Alkalaev:2009vm}.}
where $q$ is the height of the block associated with its gauge symmetry.\footnote{For unitary fields in AdS, this is the first block, so $q=h_1$.}
In the case at hand, we therefore need to consider $p$-forms $B \in \Omega^{p}(\mathrm{AdS}_{2p+1})$.
The action principle in~\cite{Alkalaev:2005kw} is of MacDowell--Mansouri--Stelle--West type~\cite{MacDowell:1977jt, Stelle:1979aj}: It is built from the linearised curvatures of an $\mathfrak{so}(2,D)$ connection, which contains the vielbein, together with a compensator --- a unit-norm vector in the fibre.
The compensator is pure gauge and carries no degrees of freedom, while the vielbein is its covariant derivative.
Our topological theories use the same type of fields, $p$-forms valued in $\mathcal{V}$, but require no compensator and have a Chern--Simons-type action.
Using the invariant bilinear form $\langle\cdot \wedge \cdot\rangle$, the dualisation operator $\spar$ defined on $\mathcal{V}$, and the Hodge dual $\star$ on the boundary, we have
\begin{equation}
    S = \int_{\text{AdS}_{2p+1}} \< \spar B \wedge \DD B \> + \frac{1}{2}\int_{\pp \text{AdS}_{2p+1}}\< B \wedge \star B\>.
    \label{eq: singleton action intro}
\end{equation}
Here $\DD$ is the AdS$_{2p+1}$-covariant derivative, which is flat in the sense that $\DD^{2}=0$.
The bulk integrand is a $(2p+1)$-form, since $B$ and $\DD B$ have degrees $p$ and $p+1$, and it is manifestly $\mathfrak{so}(2,2p)$-invariant.
The equations of motion, $\DD B=0$, set all curvatures to zero, and the gauge symmetry $\delta B = \DD A$ (which leaves them invariant since $\DD^{2}=0$) removes all local degrees of freedom.
The bulk is therefore topological, and the dynamics are pushed to the conformal boundary.
On the boundary, $B$ reduces to a $(p-1)$-form potential valued in $\mathfrak{so}(1,2p-1)$-tensors, which carries the biform Young symmetry $[p-1,p-1]$ of the singleton gauge field as expected for spin $s=2$.

\subsection*{Organisation}

This manuscript is organised as follows.
In section~\ref{sec: tqft}, we define the field content and present our bulk description of biform singletons --- this is the starting point for deriving our boundary actions.
In section~\ref{sec: cov boundary reduction}, we perform a covariant boundary reduction, which yields a manifestly Lorentz covariant action for the entire class of biform singletons.
In section~\ref{sec: non-covariant reduction}, we break manifest Lorentz covariance to reduce this boundary action to the established formulations of~\cite{Henneaux:2004jw, Joung:2012sa, Henneaux:2016opm}.
Expressions in the lightcone basis and for the covariant derivative are collected in appendix~\ref{app: lc coords}, and details of the calculations in section~\ref{sec: non-covariant reduction} are relegated to appendix~\ref{app: constraints}.

\subsection*{Conventions}

We consider $\mathcal{V}$-valued $p$-forms $B\in\Omega^p(M)$ on $M=\mathrm{AdS}_{2p+1}$, where $\mathcal{V}=\Lambda^{p+1}(\mathbb{R}^{2,2p})$ is the $(p+1)$-th exterior power of the vector representation of $\mathfrak{so}(2,2p)$, identified with its dual by the invariant metric.
In both the fibre indices $\mathcal{V}$ and the spacetime $M$, indices along $\mathbb{R}^{2,2p}$, $\mathbb{R}^{1,2p-1}$, and $\mathbb{R}^{2p-1}$ are denoted by $I,J,\dots$, $\alpha,\beta,\dots$, and $i,j,\dots$, respectively.
Fibre indices are underlined.
In the fibre, we use the lightcone basis $\ul{I} = (\ul{+},\ul{-},\ul{\alpha})$ with $\ul{\alpha} = (\ul{0},\ul{i})$.
For example, choosing orthonormal fibre vectors $V_{\ul{y}}$ and $V_{\ul{z}}$ of norms $-1$ and $+1$, respectively, we take $V_{\ul{\pm}}= \tfrac{1}{\sqrt{2}}(V_{\ul{z}}\pm V_{\ul{y}})$.
With this, the non-zero components of the flat metric are $\eta_{\ul{+-}} = 1$, $\eta_{\ul{00}} = -1$ and $\eta_{\ul{ij}} = \delta_{ij}$.

(Anti-)symmetrisation is denoted by round (square) brackets, and taken to be normalised with unit weight.
The $\mathrm{AdS}_{2p+1}$-covariant derivative is denoted by $\DD$.
The Hodge duals on the boundary $\pp M$ and on the fibre are denoted by $\star$ and $\spar$, respectively.
On a $p$-form $B$ on $\pp M$, the former acts as
\begin{equation}
    \(\star B\)_{\alpha_{1} \dots \alpha_{p}} = \frac{1}{p!}\,\epsilon^{\beta_{1} \dots \beta_{p}}{}_{\alpha_{1} \dots \alpha_{p}}\, B_{\beta_{1} \dots \beta_{p}},
\end{equation}
so that $B \wedge \star B$ is the standard norm times the volume form.
The expression for $\spar$ is given in section~\ref{ssec: field content}.
We orient $\pp M$ and $M$ by the Levi-Civita symbols, normalised as in appendix~\ref{app: lc coords}.
The choice of overall sign of these symbols is then a convention, since flipping it reverses $\star$, $\spar$, and the orientations together --- this leaves the action invariant.

We specify Young diagrams by their column lengths in square brackets $[\cdot]$. 
Repeated lengths are denoted by powers, e.g.\ $[p,p]=[p^{2}]$.

\section{Topological bulk description \label{sec: tqft}}

We will now specify the structures appearing in the topological action~\eqref{eq: singleton action intro} and present the boundary conditions that select the biform singleton sector.

\subsection{Field content \label{ssec: field content}}

The gauge fields of our bulk theory are $p$-forms $B \in \Omega^{p}(M)$ on a $(2p+1)$-dimensional manifold $M$ with boundary $\pp M$, valued in $\mathcal{V}=\Lambda^{p+1}(\mathbb{R}^{2,2p})$, the space of anti-symmetric tensors of rank $p+1$ in $\mathfrak{so}(2,2p)$.
In terms of a basis $V_{\ul{I_{1} \dots I_{p+1}}}$ of $\mathcal{V}$, we expand
\begin{equation}
    B = \frac{1}{(p+1)!}\, B^{\ul{I_{1} \dots I_{p+1}}}\, V_{\ul{I_{1} \dots I_{p+1}}},\qquad B^{\ul{I_{1} \dots I_{p+1}}} \in \Omega^{p}(M).
    \label{eq: B expansion}
\end{equation}
The space $\mathcal{V}$ carries the symmetric, $\mathfrak{so}(2,2p)$-invariant bilinear form
\begin{equation}
    \< \cdot\,\wedge \cdot \>: \mathcal{V} \times \mathcal{V} \to \mathbb{R}
\end{equation}
induced by the metric of $\mathbb{R}^{2,2p}$.
It extends to $\mathcal{V}$-valued forms: For $A\in\Omega^{m}(M)$, $B\in\Omega^{n}(M)$ and a covariant derivative $\DD$ compatible with the bilinear form (to be defined in section~\ref{ssec: singleton action}), we have
\begin{subequations}
    \begin{align}
        &\< A \wedge B \> = (-)^{mn} \< B \wedge A \> \\
        &\dd \< A \wedge B \> = \< \DD A \wedge B \> + (-)^{m} \< A \wedge \DD B \> .
    \end{align}
\end{subequations}
In terms of the components in~\eqref{eq: B expansion}, this reads
\begin{equation}
    \<A \wedge B\> = \frac{1}{(p+1)!}\, A^{\ul{I_{1} \dots I_{p+1}}} \wedge B_{\ul{I_{1} \dots I_{p+1}}}.
\end{equation}

The Levi-Civita symbol of $\mathbb{R}^{2,2p}$ defines a second Hodge dual $\spar$ on $\mathcal{V}$, which acts on the basis as
\begin{equation}
    \spar V_{\ul{I_{1} \dots I_{p+1}}} = \frac{1}{(p+1)!}\, \epsilon^{\ul{J_{1} \dots J_{p+1}}}{}_{\ul{I_{1} \dots I_{p+1}}}\,V_{\ul{J_{1} \dots J_{p+1}}}.
\end{equation}
This is compatible with the inner product in the sense that
\begin{equation}
    \< \spar A \wedge \spar B \> = \< A \wedge B \>.
\end{equation}
Explicit expressions for this are given in appendix~\ref{sapp: fibre dualisation}.
The dualisation operator $\spar$ defines a \textit{linear automorphism} of $\mathcal{V}$, satisfying
\begin{equation}
    \spar^{2}=(-)^{p+1}.
\end{equation}
Fixing a non-null vector $w^{\ul{I}}$, the space $\mathcal{V}$ splits as
\begin{equation}
    \mathcal{V} = \mathcal{W} \oplus \spar\,\mathcal{W},
    \label{eq: vector space decomposition}
\end{equation}
where $\mathcal{W}$ and $\spar \mathcal{W}$ respectively spanned by $w^{\ul{J}}\, V_{\ul{J I_{1}\dots I_{p}}}$ and $w^{\ul{J}} \,\spar V_{\ul{J I_{1} \dots I_{p}}}$.
Since $w$ is non-null, the decomposition is orthogonal, and depends on the choice of $w$.
The vector $w$ breaks $\mathfrak{so}(2,2p)$ to its stabiliser, and effectively plays the role of the \textbf{compensator} in the Stelle--West formulation~\cite{Stelle:1979aj}, as we explain below.
Accordingly, we write any $A \in \mathcal{V}$ as 
\begin{equation}
    A = A^{(1)} + \spar^{-1} A^{(2)},\qquad A^{(1)},A^{(2)} \in \mathcal{W},
    \label{eq: B decomposition}
\end{equation}
with $\spar^{-1} = (-)^{p+1}\spar$ and labels $a,b,\dots \in \{1,2\}$, so that 
\begin{equation}
	\< A \wedge B \> = \< A^{(1)} \wedge B^{(1)} \> + \< A^{(2)} \wedge B^{(2)} \>.
\end{equation}

\subsection{Bulk topological field theory \label{ssec: singleton action}}


Our central claim is that biform singletons are described by the Chern--Simons-type action
\begin{equation}
    S = \int_{M} \< \spar B \wedge \DD B \> + \frac{1}{2}\int_{\pp M}\< B \wedge \star B\>,
    \label{eq: singleton action}
\end{equation}
where $\star$ is the Hodge dual on $\pp M$ and $\DD$ is a covariant derivative compatible with the bilinear form.
The Hodge dual $\star$ in the boundary term is defined with respect to the flat metric $\eta_{\alpha\beta}$ on $\pp M = \mathbb{R}^{1,2p-1}$, with coordinates $x^{\alpha} = (t,x^{i})$.
In fact, only the conformal class of this metric matters: Under a Weyl rescaling $\eta_{\alpha\beta} \to e^{2\sigma}\,\eta_{\alpha\beta}$, the Hodge dual of a $k$-form in $2p$ dimensions picks up a factor $e^{(2p-2k)\sigma}$, which equals one for $k=p$.
The boundary term is therefore Weyl invariant, as expected for a theory whose symmetry $\mathfrak{so}(2,2p)$ acts as the conformal algebra of $\pp M$.
We leave $\DD$ still unspecified for the moment, and fix it below.

Using $\spar^{2}=(-)^{p+1}$, the Leibniz rule, and the fact that $\DD$ commutes with $\spar$, the variation of the action is
\begin{equation}
    \delta S = 2 (-)^{p+1} \int_{M} \<\delta B \wedge \spar \DD B\> +\int_{\pp M} \<\delta B \wedge \(\star B + \spar B\) \>.
\end{equation}
The bulk equation of motion sets $\DD B = 0$, while the boundary equation of motion imposes
\begin{equation}
    \star B + \spar B = 0.
    \label{eq: bdry eom}
\end{equation}
This relates the Hodge dual on the boundary to the one on the fibre, and is precisely the (twisted) self-duality relation we are after: Once $B$ is reduced to the boundary fields, it becomes the relation discussed in the introduction.

However, the unrestricted boundary spectrum is larger than the singleton sector.
To select the singleton fields, we impose the \textbf{asymptotic condition}
\begin{equation}
    v \wedge B = \mathcal{O}\(z\),
    \label{eq: asymptotic condition}
\end{equation}
where $v$ is a nowhere-null closed one-form on $M$.
These are the conditions that ensure the correct boundary degrees of freedom, analogous to the torsion-free condition in a first-order formulation of gravity, and they recast the on-shell conditions of the boundary theory as fall-off conditions in the bulk.
This is partly inspired by the fall-off conditions in the reduction of AdS$_{3}$ gravity to its conformal boundary~\cite{Brown:1986nw, Coussaert:1995zp, Donnay:2016iyk, Cotler:2018zff, Chen:2025xlo}.
We will see that they can be imposed with Lagrange multipliers without propagating additional degrees of freedom, and in section~\ref{sec: non-covariant reduction}, we show that the action supplemented by these conditions reproduces the actions of~\cite{Joung:2012sa}.

\paragraph{Covariant derivative.} Our proposal is to take 
\begin{equation}
	\DD = \dd + \Omega,\quad \Omega = \frac{1}{2}\,\Omega^{\ul{IJ}}\,J_{\ul{IJ}}
	\label{eq: cov div def}
\end{equation}
where $\Omega$ is a \textit{flat} $\mathfrak{so}(2,2p)$ connection, satisfying
\begin{equation}
    \dd \Omega + \frac{1}{2}[\Omega,\Omega] = 0,
    \label{eq: connection flatness}
\end{equation}
and $J_{\ul{IJ}}$ are the generators of $\mathfrak{so}(2,2p)$ which act on $\mathcal{V}$ as tensors.
See appendix~\ref{sapp: covariant derivatives} for more details on this.
Flatness implies that $\DD^{2}=0$, so the bulk equation of motion $\DD B=0$ is invariant under the gauge symmetry $\delta B = \DD A$ for $A \in \Omega^{p-1}(M)$.
This removes all local degrees of freedom.

Such a connection actually encodes the geometry of anti-de Sitter space, with the vielbein and spin connection combining into $\Omega$.
From now on, we will therefore take $M = \mathrm{AdS}_{2p+1}$.
In Poincar\'e coordinates $(x^{\alpha},z)$, the metric reads
\begin{equation}
    \dd s^{2} = \frac{\eta_{\alpha\beta}\,\dd x^{\alpha}\, \dd x^{\beta} + \(\dd z\)^{2}}{z^{2}}.
    \label{eq: poincare ads}
\end{equation}
The metric induced on a slice of constant $z$ is $z^{-2}\eta_{\alpha\beta}$, which diverges as $z \to 0$, so only its conformal class is defined on the conformal boundary $\pp M$ at $z=0$.
This is however all that the boundary term requires: Since $\star$ is Weyl invariant on $p$-forms, the factor $z^{-2}$ drops out, and we may use the flat representative $\eta_{\alpha\beta}$ in $\star$.
In the lightcone basis (defined in the conventions), the connection is
\begin{equation}
    \Omega = - \frac{\dd z}{z}\, J_{\ul{+-}} - \sqrt{2}\, \frac{\dd x^{\alpha}}{z}\, J_{\ul{-\alpha}}.
    \label{eq: Omega lightcone}
\end{equation}
The generator $J_{\ul{+-}}$ of dilatations grades $\mathcal{V}$ by conformal weight:
\begin{equation}
    [J_{\ul{+-}},V_{\ul{\alpha_{1} \dots \alpha_{p+1}}}] = 0,\quad [J_{\ul{+-}},V_{\ul{\pm \alpha_{1} \dots \alpha_{p}}}] = \pm V_{\ul{\pm \alpha_{1} \dots \alpha_{p}}},\quad [J_{\ul{+-}},V_{\ul{+- \alpha_{1} \dots \alpha_{p-1}}}] = 0.
    \label{eq: bracket -+}
\end{equation}
Through the $\dd z/z$ term in $\Omega$, this fixes how the components of $B$ scale near the boundary: Those with lower-index components $\ul{\pm}$ scale as $z^{\mp 1}$, and the others as $z^{0}$.

\paragraph{Action decomposition.} Since $\spar$ is $\mathfrak{so}(2,2p)$-invariant, the action splits into its eigenspaces.
This makes chirality and electric--magnetic duality explicit:
\begin{itemize}
\item For odd $p$, $\spar^{2} = 1$ with eigenvalues $\pm 1$.
The two eigenspaces $\mathcal{V}_{\pm}$ are mutually orthogonal, so the bilinear form of $\mathcal{V}$ is block-diagonal with respect to them.
Writing $B = B_{+} + B_{-}$, the action~\eqref{eq: singleton action} becomes
\begin{equation}
	S = \tilde{S}_{+}[B_{+}] - \tilde{S}_{-}[B_{-}],\quad \tilde{S}_{\pm}[B] = \int_{M} \< B \wedge \DD B \> \pm \frac{1}{2}\int_{\pp M}\< B \wedge \star B \>.
    \label{eq: chiral decomposition}
\end{equation}
Because the pairing is diagonal, the chiral halves decouple, and the action is a difference of two Chern--Simons actions with boundary terms of opposite signs.
For $p=1$, this is the action studied in~\cite{Chen:2025xlo}.
\item For even $p$, $\spar^{2} = -1$, so the eigenvalues $\pm\ii$ require the complexification $\mathcal{V}\otimes\mathbb{C}$, whose two eigenspaces are null and exchanged by complex conjugation.
The bilinear form, extended complex-bilinearly, therefore only pairs one eigenspace with the other: It is block-off-diagonal.
Since $B$ is real, $B = B_{+} + B_{-}$ with $B_{+} = \ol{B_{-}}$.
Hence, $B$ is determined by $B_{-}$ alone, and the action becomes
\begin{equation}
    S = \ii \int_{M} \< B_{+} \wedge \DD B_{-} - B_{-} \wedge \DD B_{+} \> + \int_{\pp M}\< B_{+} \wedge \star B_{-}\>.
    \label{eq: complex decomposition}
\end{equation}
Because of this off-diagonal pairing, the fields $B_{\pm}$ do not decouple, and satisfy a \textit{twisted self-duality} relation, with $\mathrm{SO}(2)$ traded for $\mathrm{U}(1)$.
\end{itemize}
The compensator decomposition~\eqref{eq: vector space decomposition} is not the eigenspace decomposition, since $\spar$ maps $\mathcal{W}$ to $\spar\,\mathcal{W}$.
Instead, in terms of the components in~\eqref{eq: B decomposition}, the eigencomponents are the combinations $(B^{(1)} \pm B^{(2)})/\sqrt{2}$ for odd $p$ and $(B^{(1)}\pm \ii B^{(2)})/\sqrt{2}$ for even $p$.
As $\<X,\spar Y\> = 0$ for $X,Y \in \mathcal{W}$,~\eqref{eq: chiral decomposition} and~\eqref{eq: complex decomposition} can thus be written entirely in terms of $\mathcal{W}$-valued fields, with $\DD$ replaced by the connection it induces on $\mathcal{W}$.

Note also that for $p=1$, this is literally the compensator formulation of AdS$_3$ gravity.
Here $\mathcal{V} = \Lambda^{2}(\mathbb{R}^{2,2}) \simeq \mathfrak{so}(2,2)$, and~\eqref{eq: singleton action} is linearised Chern--Simons gravity.
The compensator splits an $\mathfrak{so}(2,2)$ connection into a dreibein valued in $\mathcal{W}$ and a spin connection valued in $\spar\,\mathcal{W}$~\cite{MacDowell:1977jt, Stelle:1979aj}.
For $w = (V_{\ul{+}} - V_{\ul{-}})/\sqrt{2}$, which is constant in the coordinates used here, $\DD w$ is the background dreibein.
The two chiral combinations are then, up to normalisation, the $\mathfrak{sl}(2,\mathbb{R})$ connections $\omega \pm e$, and~\eqref{eq: chiral decomposition} is the familiar difference of two $\mathfrak{sl}(2,\mathbb{R})$ Chern--Simons actions.
For general $p$,~\eqref{eq: vector space decomposition} generalises this split into frame-like fields.

\section{Covariant boundary reduction \label{sec: cov boundary reduction}}

In the previous section, we presented our description of biform singletons in $D=2p$ dimensions as edge modes of topological theories in AdS$_{2p+1}$.
We will now apply a covariant boundary reduction due to~\cite{Arvanitakis:2022bnr, Evnin:2023ypu} to derive boundary action formulations which realise (twisted) self-duality while preserving manifest Lorentz covariance.

Our central result is the manifestly Lorentz-covariant action for all biform singletons,
\begin{equation}
    S_{\text{total}} = S_{\text{bdry}} + S_{\text{constr.}} + S_{v}.
    \label{eq: total action}
\end{equation}
We derive the boundary action $S_{\text{bdry}}$ in section~\ref{ssec: cov boundary action}, the constraint term $S_{\text{constr.}}$ imposing the asymptotic conditions \eqref{eq: asymptotic condition} in section~\ref{ssec: asymptotic}, and the term $S_{v}$ ensuring manifest Lorentz covariance in section~\ref{ssec: lorentz}.

\subsection{Boundary reduction \label{ssec: cov boundary action}}

Let us begin by reducing the topological action \eqref{eq: singleton action} to the boundary.
As an ansatz, we follow the prescription in \cite{Arvanitakis:2022bnr}, and take
\begin{equation}
    B =  C + v \wedge \tilde{R},
    \label{eq: reduction ansatz}
\end{equation}
where $v$ is the nowhere-null closed one-form of~\eqref{eq: asymptotic condition}, $\tilde{R} \in \Omega^{p-1}(M)$ is $\mathcal{V}$-valued, and $C \in \Omega^{p}(M)$ is the remaining part of $B$.
This split is not unique, but the redundancy drops out in the following.
Substituting this into the first (bulk) term of the action~\eqref{eq: singleton action} and integrating by parts, 
\begin{equation}
    \int_{M} \< \spar B \wedge \DD B \> = \int_{M} \< \spar C \wedge \DD C + 2  (-)^{p+1}\, v \wedge \tilde{R} \wedge \spar \DD C \>+ \int_{\pp M}  \<v \wedge \tilde{R} \wedge \spar C\>.
\end{equation}
The field $\tilde{R}$ acts as a Lagrange multiplier, enforcing the condition
\begin{equation}
    v \wedge \DD C = 0
\end{equation}
in the bulk.
Since the operator $\DD$ is nilpotent, this is locally solved by
\begin{equation}
    C = \DD A + v \wedge P,
\end{equation}
where $A,P \in \Omega^{p-1}(M)$.
This amounts to taking our ansatz~\eqref{eq: reduction ansatz} to be
\begin{equation}
    B = H \equiv \DD A + v \wedge R,\quad R = \tilde{R} + P,
\end{equation}
although for convenience, we will often denote $F \equiv \DD A$ --- in other words, from now on it is understood that $F$ is $\DD$-exact.
Substituting this solution of the bulk constraint into the action \eqref{eq: singleton action} reduces it to the following boundary integral:
\begin{equation}
    \begin{aligned}
        S_{\text{bdry}} &= \frac{1}{2} \int_{\pp M} \< H\wedge \star H + 2 v \wedge R \wedge \spar F \> \\
 &= \frac{1}{2} \int_{\pp M} \<F \wedge \star F + 2v \wedge R \wedge \(\spar F + \star F \) + v \wedge R \wedge \star \(v \wedge R \)\>.
    \end{aligned}
    \label{eq: singleton action 2}
\end{equation}
This is the first ingredient in the total action \eqref{eq: total action}.

\paragraph{Dynamical content.} Note that~\eqref{eq: singleton action 2} has a large set of gauge symmetries:
\begin{itemize}
    \item $\DD$-exact shifts in $A$: $\delta A = \DD \alpha$,
    \item $v$-exact shifts in $R$: $\delta R = v\wedge \rho $,
    \item 
    combined shifts $\delta A = v \wedge \lambda$, $\delta R = \DD \lambda$ with $\lambda \in \Omega^{p-2}(M)$, which leave $H$ invariant.
\end{itemize}
Next, note the equation of motion derived from the auxiliary field $R$ is
\begin{equation}
    v \wedge \(\spar H + \star H\) = 0.
    \label{eq: H twisted SD}
\end{equation}
The combination $\mathcal{E} = \spar H + \star H$ is manifestly twisted self-dual, i.e. $\star \mathcal{E} = \spar \mathcal{E} $.
Since $v \wedge$ commutes with $\spar$, $v\wedge \mathcal{E} = 0$ implies $v \wedge \spar \mathcal{E} = v \wedge \star \mathcal{E} = 0$.
This forces $\mathcal{E} = 0$ for non-null $v$ --- see~\cite{Pasti:2012wv, Bansal:2021bis} for more details.
The equation for $A$ is $\DD (\star H - \spar H) =0$.
Using $\mathcal{E} = 0$, we then find
\begin{equation}
    v \wedge \DD R =0.
\end{equation}
Locally, this is solved by $R = \DD \lambda + v \wedge \rho$, which can be removed by the gauge symmetries above.
In other words, $R$ is pure gauge.
Now, by fixing the gauge symmetry in $R$, we get
\begin{equation}
    \star \DD A + \spar \DD A = 0.
\end{equation}
This takes the desired form of a twisted self-duality relation~\eqref{eq: bdry eom}.

\subsection{Asymptotic conditions \label{ssec: asymptotic}}

Since $v \wedge H = v \wedge F$, the asymptotic condition~\eqref{eq: asymptotic condition} becomes
\begin{equation}
    v \wedge F = \mathcal{O}(z).
\end{equation}
The explicit expressions for the field strength $F$ are given in~\eqref{eq: field strength components} in appendix~\ref{sapp: covariant derivatives}.

To analyse this condition, we expand the components of $A$ near $z=0$ according to their conformal weight, as anticipated in section~\ref{ssec: singleton action}:
\begin{subequations}
    \begin{align}
        A_{\ul{- \alpha_{2} \dots \alpha_{p+1}}} &= z^{+1}\, \hat{A}_{\ul{- \alpha_{2} \dots \alpha_{p+1}}}+ \dots \\
        A_{\ul{\alpha_{1} \dots \alpha_{p+1}}} &= \frac{1}{\sqrt{2}}\,z^{0}\, \hat{A}_{\ul{\alpha_{1} \dots \alpha_{p+1}}} + \dots \\
        A_{\ul{-+\alpha_{3} \dots \alpha_{p+1}}} &= \frac{1}{\sqrt{2}}\,z^{0}\, \hat{A}_{\ul{-+ \alpha_{3} \dots \alpha_{p+1}}} + \dots \\
        A_{\ul{+ \alpha_{2} \dots \alpha_{p+1}}} &= z^{-1}\, \hat{A}_{\ul{+ \alpha_{2} \dots \alpha_{p+1}}} + \dots\,.
     \end{align}
     \label{eq: field rescalings}%
\end{subequations}
Note in particular that the fields on the right-hand side are coefficients of the leading term in the asymptotic expansion near $z = 0$ and are independent of $z$.
The powers of $z$ follow from the $J_{\ul{+-}}$ eigenvalues~\eqref{eq: bracket -+}.
In terms of this, the field strength components~\eqref{eq: field strength components} read
\begin{subequations}
    \begin{align}
        z^{+1}\, F_{\ul{+ \alpha_{1} \dots \alpha_{p}}} &= \dd \hat{A}_{\ul{+ \alpha_{1} \dots \alpha_{p}}}  - \dd x^{\beta}\wedge \hat{A}_{\ul{\beta \alpha_{1} \dots \alpha_{p}}} - p\, \dd x_{[\alpha_{1}|} \wedge \hat{A}_{\ul{-+|\alpha_{2} \dots \alpha_{p}]}} + \dots  \\
        \sqrt{2}F_{\ul{\alpha_{1}\dots \alpha_{p+1}}} &= \dd \hat{A}_{\ul{\alpha_{1}\dots \alpha_{p+1}}} +2(p+1) \dd x_{[\alpha_{1}|} \wedge \hat{A}_{\ul{-|\alpha_{2}\dots \alpha_{p+1}]}} + \dots \\
        \sqrt{2}F_{\ul{+- \alpha_{1} \dots \alpha_{p-1}}} &= \dd \hat{A}_{\ul{+-\alpha_{1}\dots \alpha_{p-1}}} + 2 \dd x^{\beta} \wedge \hat{A}_{\ul{- \beta \alpha_{1} \dots \alpha_{p-1}}} + \dots \\
        z^{-1}\, F_{\ul{- \alpha_{1} \dots \alpha_{p}}} &= \dd  \hat{A}_{\ul{- \alpha_{1} \dots \alpha_{p}}}  + \dots,
    \end{align}
\end{subequations}
where, from now on, $\DD$ acting on boundary fields is understood to be in the sense above.
The components $F_{\ul{-\alpha_{1}\dots\alpha_{p}}}$ are of order $z$ and satisfy~\eqref{eq: asymptotic condition} automatically.
The remaining components, $F_{\ul{+\alpha_{1}\dots\alpha_{p}}}$, $F_{\ul{-+\alpha_{1}\dots\alpha_{p-1}}}$ and $F_{\ul{\alpha_{1}\dots\alpha_{p+1}}}$, start at order $z^{-1}$ or $z^{0}$.
On the leading coefficients, the conditions~\eqref{eq: asymptotic condition} therefore read
\begin{equation}
    v\wedge \hat{F}_{\ul{+\alpha_{1} \dots \alpha_{p}}} = 0,\quad v\wedge \hat{F}_{\ul{-+\alpha_{1} \dots \alpha_{p-1}}} = 0,\quad
    v\wedge \hat{F}_{\ul{\alpha_{1} \dots \alpha_{p+1}}} = 0,
    \label{eq: on shell conditions}
\end{equation}
where $\hat{F}$ denotes the right-hand sides above. In contrast, $\hat{F}_{\ul{-\alpha_{1}\dots\alpha_{p}}}$ is left unconstrained.
For $F_{\ul{+\alpha_{1}\dots\alpha_{p}}}$, \eqref{eq: asymptotic condition} also restricts subleading coefficients.
These drop out of the boundary action: Here, $F_{\ul{+\alpha_{1}\dots\alpha_{p}}}$ only pairs with $F_{\ul{-\alpha_{1}\dots\alpha_{p}}}$, through $\eta^{\ul{+-}}=1$, and this product is finite at $z=0$ only through the leading terms.

We can now impose these asymptotic conditions directly in the boundary action via a $\mathcal{V}$-valued Lagrange multiplier $\chi \in \Omega^{p-1}(\pp M)$:
\begin{equation}
    S_{\text{constr.}} = \int_{\pp M} \<v \wedge \chi \wedge \hat{F}\>,
    \label{eq: action constraint}
\end{equation}
with $\chi^{\ul{-\alpha_{1} \dots \alpha_{p}}} = 0$ identically, so that $\hat{F}_{\ul{-\alpha_{1}\dots\alpha_{p}}}$ is left unconstrained.
This is the second ingredient in \eqref{eq: total action}.
From now on, we work with these boundary fields and drop the hats.

\subsection{Lorentz covariance \label{ssec: lorentz}}

We claimed that the theory \eqref{eq: total action} enjoys manifest Lorentz covariance.

For this to hold, $v$ must be a dynamical field rather than a fixed one-form: a fixed $v$ selects a preferred direction on $\pp M$, and hence breaks Lorentz symmetry.
Therefore, closedness $\dd v=0$ must arise as a dynamical equation of motion.
This can be achieved by an additional Lagrange multiplier $\mu$ which is an ordinary $(2p-2)$-form \cite{Evnin:2023ypu}:
\begin{equation}
    S_{v} = \int_{\pp M}\, \mu \wedge \dd v.
    \label{eq: action v}
\end{equation}
This is the final ingredient in \eqref{eq: total action}.

Varying $\mu$ imposes $\dd v = 0$, so that locally $v = \dd a$.
For~\eqref{eq: total action} to describe the same degrees of freedom, $a$ must not propagate.
We show below that the total action is invariant under shifts $v \to v + \dd\varphi$ (as long as $v$ stays nowhere-null), so that $v$ is pure gauge.
The argument is somewhat subtle.

\paragraph{Constraint term.} We first show that adding $S_{\text{constr.}}$ does not modify the singleton sector.
The multiplier $\chi$ pairs with all components of $F$ except $F_{\ul{-\alpha_{1}\dots\alpha_{p}}}$.
We denote these components by a subscript $\parallel$, so that, for instance, $A_{\parallel}$ collectively denotes $A_{\ul{\alpha_{1} \dots \alpha_{p+1}}}$, $A_{\ul{+\alpha_{1} \dots \alpha_{p}}}$ and $A_{\ul{-+\alpha_{1} \dots \alpha_{p-1}}}$.
For the components $F_{\ul{-\alpha_{1}\dots\alpha_{p}}}$, the previous discussion proceeds unchanged.
The equation of motion for $R$ remains~\eqref{eq: H twisted SD}, and the equations for $\chi$ are precisely the constraints themselves:
\begin{equation}
    v \wedge H_{\parallel} = 0.
\end{equation}
Since $\spar$ maps the $\parallel$ components among themselves (cf. appendix~\ref{app: lc coords}), the fibre dual of this equation also holds.
Together with  \eqref{eq: H twisted SD}, we then arrive at
\begin{equation}
    v \wedge \star H_{\parallel} = 0.
\end{equation}
Once again,  since $v$ is non-null, this implies $H_{\parallel} = 0$.
From this, it follows that $v \wedge \DD R_{\parallel} = 0$, and  $R_{\parallel}$ is pure gauge.

Next, note that the equation for $A_{\parallel}$ is
\begin{equation}
    \DD\(\star H_{\parallel} - \spar H_{\parallel} + v\wedge \chi\)=0.
\end{equation}
However, we just showed that $H_{\parallel}=0$, so
\begin{equation}
    v\wedge \DD\chi=0.
    \label{eq: chi constraint}
\end{equation}
Locally, this is solved by $\chi = \DD\alpha + v \wedge \beta$, which is a gauge transformation of~\eqref{eq: singleton action 2} together with~\eqref{eq: action constraint}, so $\chi$ is pure gauge.

\paragraph{Shift symmetry of $v$.} We can now show that $v$ is pure gauge.
Writing $v = \dd a$, the term $S_{v}$ drops out of the equation for $a$, which reads
\begin{equation}
    \dd\left\langle R \wedge (\star H + \spar H) - \chi \wedge H\right\rangle =0.
\end{equation}
The first term vanishes by $\mathcal{E} = 0$, and the second because $\chi$ pairs only with $H_{\parallel} = 0$; this is where $H_{\parallel} = 0$ is crucial.
The equation for $a$ is therefore implied by the others.
Correspondingly, shifts $a \to a + \varphi$, accompanied by suitable transformations of the other fields, are a gauge symmetry, and $v$ is pure gauge, as anticipated.

\section{Non-covariant boundary reduction \label{sec: non-covariant reduction}}


So far, we have presented two formulations of biform singletons,~\eqref{eq: singleton action} and~\eqref{eq: total action}.
To confirm that they describe biform singletons, we now connect them to the actions of~\cite{Joung:2012sa}, which for $p=2$ reduce to the formulation of four-dimensional linearised gravity in~\cite{Henneaux:2004jw}.
This requires fixing the gauge, eliminating $R$, and solving the constraints imposed by $\chi$.
This breaks manifest Lorentz covariance.

Since $v$ is pure gauge, we may choose $v=\dd t$.
This splits the boundary as $\pp M=\mathbb{R}_t\times\Sigma$, where $\Sigma\simeq\mathbb{R}^{2p-1}$ has coordinates $x^{i}$.
In the fibre, we correspondingly take the compensator along the time direction, $w = V_{\ul{0}}$.

\subsection{Hamiltonian form \label{ssec: gauge}}

Using the shifts $\delta R=v\wedge\rho$ and $\delta A=v\wedge\lambda$, we set
\begin{equation}
    \iota_{\pp_t}\,R = 0,\quad \iota_{\pp_t}\,A = 0,
    \label{eq: gauge fixing}
\end{equation}
so that $A$ and $R$ become $t$-dependent forms on $\Sigma$.
The field strength then splits as
\begin{equation}
    F=\dd t\wedge F_{t}+F_\Sigma,
\end{equation}
with $F_{t} \equiv\DD_t A$ and $F_\Sigma\equiv\DD_\Sigma A$.

Let us denote the Hodge dual on $\Sigma$ as $*$, so that $*^2=1$, and take $\epsilon_{0i_1\dots i_{2p-1}}=\epsilon_{i_1\dots i_{2p-1}}$ (cf. appendix~\ref{app: lc coords}).
Then the Hodge dual on $\pp M$ splits as
\begin{equation}
    \star(\dd t\wedge\alpha+\beta)=-*\alpha+(-)^p\,\dd t\wedge*\beta,
\end{equation}
given $\alpha\in\Omega^{p-1}(\Sigma)$ and $\beta\in\Omega^{p}(\Sigma)$.
The boundary action~\eqref{eq: singleton action 2} becomes
\begin{equation}
   S_{\text{bdry}}=\int_{\pp M}\,\dd t \wedge \<\frac{1}{2} \, F_{\Sigma} \wedge * F_{\Sigma}-\frac{1}{2} \, (F_{t}+R) \wedge * (F_{t}+R) + R \wedge \spar F_{\Sigma}\>.
   \label{eq: singleton action 3}
\end{equation}
The equation of motion for the auxiliary field $R$ is $	*(F_{t}+R)=\spar F_\Sigma$ so after substituting this back gives
\begin{equation}
   S= \int_{\pp M} \, \dd t \wedge \< F_{\Sigma} \wedge* F_{\Sigma}- F_{t} \wedge \spar  F_{\Sigma} \>.
   \label{eq: singleton action 4}
\end{equation}
This is first-order in time, with $\spar$ acting as a duality rotation, and extremely reminiscent of the Schwarz--Sen action~\cite{Schwarz:1993vs}.
The auxiliary field $R$ no longer appears below, so $R$ will be reserved to denote a curvature.

\subsection{Conformal Cartan geometry \label{ssec: cartan}}

With $v=\dd t$, the asymptotic conditions~\eqref{eq: on shell conditions} restrict the spatial field strength $F_{\Sigma}$ to the component $F_{\Sigma\,\ul{-\alpha_{1}\dots\alpha_{p}}}$ only.
Since $\eta^{\ul{--}}=0$ and the dualisation $\spar$ preserves this type of component, the first term of~\eqref{eq: singleton action 4} drops out
\begin{equation}
   S=-\int_{\pp M}\,\dd t\wedge\<F_{t} \wedge\spar F_\Sigma\>.
\end{equation}
In this, only $F_{t\ul{+\alpha_{1}\dots\alpha_{p}}}$ contributes.

We now use the compensator $w = V_{\ul{0}}$ in~\eqref{eq: vector space decomposition}.
Then $\mathcal{W}\simeq\Lambda^{p}(\mathbb{R}^{1,2p})$, where $\mathbb{R}^{1,2p}=\mathrm{span}\{V_{\ul{+}},V_{\ul{-}},V_{\ul{i}}\}$ carries the conformal algebra $\mathfrak{so}(1,2p)$ of $\Sigma$.
In terms of the components~\eqref{eq: B decomposition}, the action reads
\begin{equation}
   S=-\int_{\pp M}\,\dd t\wedge\hat\epsilon_{ab}\,\<F_{t}^{(a)}\wedge F^{(b)}_\Sigma\>,
   \label{eq: singleton action 5}
\end{equation}
where $\hat\epsilon_{12}=1$ and $\hat\epsilon_{21}=(-)^{p+1}$, i.e.  $\hat{\epsilon}$ is $(-)^{p+1}$-symmetric.
The dilatation generator $J_{\ul{+-}}$ defines the grading
\begin{equation}
	\mathbb{R}^{1,2p} =\mathrm{span}\{V_{\ul{+}}\}\oplus\mathrm{span}\{V_{\ul{i}}\}\oplus\mathrm{span}\{V_{\ul{-}}\},
\end{equation}
with respective weights $+1,0,-1$.
This grading reflects the parabolic structure of conformal geometry on $\Sigma$ and induces a decomposition of $\mathcal{W}$ into three grades.
Accordingly, we name the components of $A^{(a)}$ after the conformal generators of $\Sigma$:
\begin{equation}
    \begin{gathered}
        Z^{(a)}_{\ul{i_{1} \dots i_{p-1}}} \equiv A^{(a)}_{\ul{+ i_{1} \dots i_{p-1}}},\quad S^{(a)}_{\ul{i_{1} \dots i_{p-1}}} \equiv A^{(a)}_{\ul{- i_{1} \dots i_{p-1}}}, \\
        \omega^{(a)}_{\ul{i_{1} \dots i_{p}}} \equiv A^{(a)}_{\ul{ i_{1} \dots i_{p}}},\quad \phi^{(a)}_{\ul{i_{1} \dots i_{p-2}}} \equiv A^{(a)}_{\ul{-+ i_{1} \dots i_{p-2}}}.
    \end{gathered}
    \label{eq: gauge fields}
\end{equation}
Here, components of $\mathcal{W}$-valued fields are labelled by their $p$ indices other than $\ul{0}$.
They generalise translations, special conformal transformations, rotations, and dilatations, respectively.
Each is a $(p-1)$-form on $\Sigma$ valued in antisymmetric tensors of $\mathfrak{so}(2p-1)$, i.e. a biform on $\Sigma$.

Note in particular that for $p=2$, $\Lambda^2(\mathbb{R}^{1,4})\simeq\mathfrak{so}(1,4)$ is the adjoint representation.
$A^{(a)}$ is then a linearised Cartan connection for the conformal geometry of the three-dimensional slices, and $\DD$ acts on it through the flat Cartan connection.
In this language, the asymptotic conditions require all components of the curvature except $F_{\ul{-\alpha_{1}\dots\alpha_{p}}}$ to vanish, so that the curvature lies in a single grade.
In three dimensions, this is precisely the normality condition of a conformal Cartan connection, whose only remaining curvature is the Cotton tensor.
For general $p$, $\Lambda^p(\mathbb{R}^{1,2p})$ replaces the adjoint, and the same structure holds with biforms in place of tensors.

\subsection{Solving the constraints \label{ssec: constraints}}

We will now solve the asymptotic conditions.
We treat all fields as forms on $\Sigma$ that carry antisymmetric fibre indices.
The spatial analysis below applies to each component $A^{(a)}$ separately, so we suppress the label $(a)$ until the two components mix.

Let $\bar{e}^{\,i}=\dd x^i$ be the background vielbein of $\Sigma$.
For a field $X$ with $m$ fibre indices, we define two algebraic operations:
\begin{equation}
    (\bar{e}\lrcorner X)_{\ul{i_1\dots i_{m-1}}}\equiv\bar{e}^{\,j}\wedge X_{\ul{j\,i_1\dots i_{m-1}}},\qquad (\bar{e}\wedge X)_{\ul{i_1\dots i_{m+1}}}\equiv(m+1)\,\bar{e}_{[\ul{i_1}}\wedge X_{\ul{i_2\dots i_{m+1}}]}.
\end{equation}
Both operations increase the form degree by one, although $\bar{e}\lrcorner$ lowers the number of fibre indices by one while $\bar{e}\wedge$ raises it by one.
The first moves a fibre index into the form degree, and the second converts a form index into a fibre index.
With these operations, the remaining gauge symmetry $\delta A=\DD_\Sigma\Lambda$, with $\Lambda\in\Omega^{p-2}(\Sigma)$ and components $(\xi,\rho,\sigma,\kappa)$ labelled like $(Z,\omega,\phi,S)$, reads
\begin{equation}
    \delta Z=\dd\xi-\bar{e}\lrcorner\rho-\bar{e}\wedge\sigma,\qquad \delta\omega=\dd\rho+2\,\bar{e}\wedge\kappa,\qquad \delta\phi=\dd\sigma-2\,\bar{e}\lrcorner\kappa,\qquad \delta S=\dd\kappa.
\end{equation}
This is the structure familiar from conformal gravity.
For $p=2$, $Z_{\ul{i}}$ is the fluctuation of the dreibein, $\omega_{\ul{ij}}$ of the spin connection, $\phi$ the dilatation gauge field, and $S_{\ul{i}}$ the special conformal gauge field.
The parameters $\xi,\rho,\sigma,\kappa$ are then infinitesimal translations, Lorentz rotations, Weyl rescalings and special conformal transformations.

In conformal gravity ($p=2$), $\kappa$ is used to remove the dilatation gauge field --- this is possible since $\bar{e}\lrcorner\kappa=\kappa_j\,\bar{e}^{\,j}$ is an arbitrary one-form.
Then, $\rho$ is used to remove the antisymmetric part of the dreibein fluctuation, leaving a symmetric metric fluctuation $h_{ij}=Z_{(\ul{i}j)}$.
This has residual gauge transformations $\delta h_{ij}=\pp_{(i}\,\xi_{j)}-\sigma\,\delta_{ij}$, which correspond to linearised diffeomorphisms and Weyl rescalings.
This generalises to general $p$.
The shift $\bar{e}\lrcorner\kappa$ covers all of $\phi$, so we may set $\phi=0$.
The shift $\bar{e}\lrcorner\rho$ removes every part of $Z$ except the one with Young symmetry $[p-1,p-1]$,\footnote{Two sets of $p-1$ antisymmetric indices decompose into two-column Young diagrams $[p-1+k,\,p-1-k]$ with $k=0,\dots,p-1$.
The shift by $\bar{e}\lrcorner\rho$, with $p$ and $p-2$ antisymmetric indices, covers exactly the diagrams with $k\geq1$, which leaves only $[p-1,p-1]$.}
which generalises the symmetric metric fluctuation.
Denoting the projection onto this symmetry-type by $\Pi$, the residual gauge symmetry is
\begin{equation}
    \delta Z=\Pi\,\dd\xi-\bar{e}\wedge\sigma,
    \label{eq: gauge transf Z}
\end{equation}
which generalises linearised diffeomorphisms ($\xi$) and Weyl rescalings ($\sigma$) on $\Sigma$.

By the same argument, the spatial field strength $F_\Sigma=\DD_\Sigma A$ has the components
\begin{equation}
    F_{\ul{+}}=\dd Z-\bar{e}\lrcorner\omega-\bar{e}\wedge\phi,\qquad F_{\ul{-+}}=\dd\phi-2\,\bar{e}\lrcorner S,\qquad F_{\ul{\,\cdot\,}}=\dd\omega+2\,\bar{e}\wedge S,\qquad F_{\ul{-}}=\dd S,
\end{equation}
where spatial fibre indices on $\Sigma$ are suppressed, and $F_{\ul{\,\cdot\,}}$ denotes the component whose fibre indices are all spatial.
The asymptotic conditions set the first three to zero and leave $F_{\ul{-}}$ free.
We solve them on by one, mirroring the standard constraints in conformal gravity.

\textit{Torsion.} In the gauge $\phi=0$, the condition $F_{\ul{+}}=0$ reads $\dd Z=\bar{e}\lrcorner\omega$ --- this determines $\omega$ in terms of $Z$.
Since $Z$ has symmetry $[p-1,p-1]$, the only non-zero projection of $\dd Z$ has symmetry $[p,p-1]$, and so does $\omega$.
Using an identity of mixed-symmetry tensors derived in appendix~\ref{app: bianchi identity identity}, we find
\begin{equation}
    \omega=(\dd Z)^{\top},
    \label{eq: p constraint}
\end{equation}
where ${}^\top$ exchanges the fibre and form indices.
For $p=2$, this is the torsion constraint that expresses the spin connection in terms of the dreibein.

\textit{Dilatation curvature.}
With $\phi=0$, the condition $F_{\ul{-+}}=0$ becomes the algebraic constraint $\bar{e}\lrcorner S=0$.
As for $Z$, it forces $S$ to have symmetry $[p-1,p-1]$.
For $p=2$, this is the statement that the special conformal gauge field is symmetric.

\textit{Rotation curvature.}
Finally, $F_{\ul{\,\cdot\,}}=0$ reads
\begin{equation}
    R + 2\,\bar{e}\wedge S = 0,
    \label{eq: schouten equation}
\end{equation}
where $R \equiv \dd\omega$ is the linearised curvature built from $Z$ through~\eqref{eq: p constraint}.
For $p=2$, $R$ is the linearised Riemann tensor of $\Sigma$.
Both terms carry $p$ antisymmetric form indices and $p$ antisymmetric fibre indices.
Since no Young diagram of $\mathrm{SO}(2p-1)$ has two columns of height $p$, $R$ is pure trace.
This generalises the vanishing of the Weyl tensor in three dimensions.
As a consequence, $R$ is fully determined by its double dual
\begin{equation}
    G \equiv *\,\ul{*}\,R,
\end{equation}
where $*$ and $\ul{*}$ are the Hodge duals on the form and fibre indices on $\Sigma$, respectively.
$G$ carries $p-1$ form and $p-1$ fibre indices, and is the generalised Einstein tensor of $Z$~\cite{Joung:2012sa}.
For $p=2$, it is literally the linearised Einstein tensor.
Applying the double dual to~\eqref{eq: schouten equation} turns $\bar{e}\wedge$ into a contraction with the generalised Kronecker delta, which is invertible on fields with $p-1$ form and $p-1$ fibre indices.
Denoting this by $\gamma^{-1}$ (cf. \eqref{eq: def gamma inv}) as in~\cite{Joung:2012sa}, we obtain
\begin{equation}
    S = -\frac{1}{2}\,\gamma\,G,
    \label{eq: schouten solution}
\end{equation}
which is the generalised Schouten tensor of $Z$.
Once again, for $p=2$, it is the linearised Schouten tensor.
This reproduces the conformal gravity statement that, in three dimensions, the rotation curvature vanishes once the special conformal gauge field equals the Schouten tensor.
The explicit forms of $\gamma^{-1}$ and $\gamma$ are given in appendix~\ref{app: schouten identity}.

\paragraph{Final action.} Having solved the constraints, only $Z$ remains, with $\omega$ and $S$ given by~\eqref{eq: p constraint} and~\eqref{eq: schouten solution}.
The remaining spatial field strength is
\begin{equation}
    C \equiv F_{\ul{-}} = \dd S,
\end{equation}
which for $p=2$ is the linearised Cotton tensor.

The time component $F_{t}$ contains, besides $\pp_t A$, the action of the time component of $\Omega$ on $A$.
This term mixes the two components $A^{(a)}$, so we now restore the label $(a)$.
The component of $F_{t}^{(a)}$ along $\ul{+}$ then reads
\begin{equation}
    (F_{t}^{(a)})_{\ul{+}} = \pp_t Z^{(a)} + \hat{\epsilon}_{ab}\,\ul{*}\,\omega^{(b)}.
\end{equation}
Since only these components of $F_{t}^{(a)}$ pair with $F_\Sigma$, and $\hat{\epsilon}_{ab}\,\hat{\epsilon}_{ac} = \delta_{bc}$, the action becomes
\begin{equation}
   S = -\int_{\pp M}\,\dd t\wedge\Big[\hat{\epsilon}_{ab}\,\<\pp_t Z^{(a)} \wedge C^{(b)}\> + \delta_{ab}\,\<\ul{*}\,\omega^{(a)} \wedge C^{(b)}\>\Big].
\end{equation}
Integrating the second term by parts on $\Sigma$, with $\dd\,\ul{*}\,\omega = \ul{*}\,R$, we arrive at
\begin{equation}
   S = -\int_{\pp M}\,\dd t\wedge\Big[\hat{\epsilon}_{ab}\,\<\pp_t Z^{(a)} \wedge \dd S^{(b)}\> + (-)^{p}\,\delta_{ab}\,\<S^{(a)} \wedge \ul{*}\,R^{(b)}\>\Big].
   \label{eq: singleton action 6}
\end{equation}
The first term is first-order in time and pairs the two fields through the generalised Cotton tensor.
In the second term, $\ul{*}R$ is the Hodge dual of the generalised Einstein tensor $G$, so that, with~\eqref{eq: schouten solution}, it is a generalised Einstein--Schouten term, quartic in derivatives of $Z^{(a)}$.

Up to normalisation, this recovers (3.32) of~\cite{Joung:2012sa} upon replacing $p\to p-1$ and the spatial direction $z$ by the time $t$, and identifying $Z^{(1)}$ and $Z^{(2)}$ with the two fields $\mathcal{V}$ and $\mathcal{U}$ there.
The two contributions to the first term of~\eqref{eq: singleton action 6} are equal after integration by parts, since the map from $Z$ to $\dd S$ is self-adjoint with respect to $\<\cdot\,,\cdot\>$ up to the sign $(-)^{p+1}$.
This compensates that of $\hat{\epsilon}_{ab}$ (cf.\ (3.39) of~\cite{Joung:2012sa}); this assignment is therefore a matter of convention.
There, the action arises by solving the Hamiltonian constraints of a massless action with respect to $z$; here, the same constraints arise as the grade-by-grade asymptotic conditions on the conformal Cartan connection of $\Sigma$.

For odd $p$, the action \eqref{eq: singleton action 6} decomposes into independent chiral sectors.
In particular, for $p=3$, the chiral sectors are of the same form as the prepotential actions for chiral $(2,2)$-tensors in $D=6$~\cite{Henneaux:2016opm, Henneaux:2017xsb}, where the extension to general two-column chiral tensors was also anticipated.
For even $p$, it is instead manifestly invariant under $\mathrm{SO}(2)$ rotations of the two prepotentials $Z^{(a)}$.
Here, for $p=2$, this reduces to the well-known formulation of four-dimensional linearised gravity in~\cite{Henneaux:2004jw}, whose gauge invariances under linearised diffeomorphisms and Weyl rescalings of the prepotentials were made manifest in~\cite{Bunster:2012km}.
In their language, the fields $Z^{(a)}$ are the so-called superpotentials, and the residual gauge symmetry~\eqref{eq: gauge transf Z} consists of linearised diffeomorphisms and Weyl rescalings of the three-dimensional slices.
Our reduction of the covariant action~\eqref{eq: singleton action 2} to~\eqref{eq: singleton action 6} thus explains the origin of their observation that the action for linearised gravity is invariant under these.
In the Cartan picture of section~\ref{ssec: cartan}, $Z^{(a)}$ and $\omega^{(a)}$ are the linearised dreibein and spin connection on $\Sigma$,~\eqref{eq: p constraint} is the torsion constraint,~\eqref{eq: schouten solution} identifies $S^{(a)}$ with the linearised Schouten tensor, and $C^{(a)}$ and $\ul{*}R^{(a)}$ are the linearised Cotton and Einstein tensors.
For general $p$, the same construction involves the corresponding mixed-symmetry tensors, and the second term in~\eqref{eq: singleton action 6} is related to a Weyl-squared term up to Gauss--Bonnet terms.

\section{Concluding remarks \label{sec: concluding remarks}}

In this work, we presented topological field theories in AdS$_{2p+1}$ which propagate biform singletons in their conformal boundary $\mathbb{R}^{1,2p-1}$ as edge modes.
The biform gauge fields are realised as $p$-forms on AdS$_{2p+1}$ valued in a non-trivial module of its isometry algebra.
Our results are threefold.
First, we constructed a Chern--Simons-type action~\eqref{eq: singleton action} based on a flat $\mathfrak{so}(2,2p)$ connection, whose boundary equation of motion is a (twisted) self-duality relation.
Second, by a covariant boundary reduction with a closed one-form $v$ that is rendered pure gauge, we obtained boundary actions~\eqref{eq: total action} that are manifestly Lorentz covariant.
Third, breaking Lorentz covariance, we recovered the actions of~\cite{Joung:2012sa}, and found that the asymptotic conditions take the form of the normality conditions of a conformal Cartan connection on the spatial slices.

\hspace{2em} Several future directions are evident.
As we outlined in the introduction, construction of topological bulk actions that support conformal edge modes is expected to work for a large class of fields with rectangular Young diagrams, as shown in \ref{fig: window}.
The generalisation to $s>2$ is more involved, as the module $\mathcal{V}$ is no longer an exterior power and the analogue of the Cartan-geometric structure of section~\ref{ssec: cartan} remains to be spelt out. 
We also expect this to apply to more general, non-unitary conformal fields \cite{Berstein:1971, Shaynkman:2004vu, Barnich:2015tma, Bourget:2017kik}.

While we have been able to identify the field content of the topological actions \eqref{eq: singleton action} ultimately by performing a non-covariant boundary reduction to recover the formulations in \cite{Henneaux:2004jw, Joung:2012sa}, it would be instructive to obtain the standard manifestly Lorentz covariant, two-derivative action formulations for biform singletons (e.g. the Fierz--Pauli action in the case of $p=2$) directly as well.

In performing our boundary reduction in Poincar\'e coordinates, we land on a boundary theory on $2p$-dimensional flat space. 
However, as alluded to below \eqref{eq: singleton action}, the boundary theory is only sensitive to the conformal class of the boundary metric. 
We can therefore equivalently put the boundary theories on $2p$-dimensional de Sitter or Anti-de Sitter backgrounds. 
This can be done by, instead of Poincar\'e coordinates, choosing charts in which the slices are locally (A)dS.
The resulting theories are equivalent: In this sense, the theories are Weyl invariant.
This is a property shared by all singleton theories.
For $s=1$, this is reflected in the fact that the theories realise conformal symmetry in an ordinary, local manner.
For $s \geq 2$, this is no longer the case.
However, for $p=1$ with arbitrary $s$ and $(s,p)=(2,2)$, it was noted that correlation functions of gauge-invariant local operators are identical to those fixed by global conformal invariance \cite{Farnsworth:2021zgj, Farnsworth:2024yeh} --- for half-integer spin, the same was found for $p=1$~\cite{Chen:2026nws}.
These theories were thus said to enjoy `hidden conformal symmetry'.
In \cite{hifj}, the case is made that for $p=1$ this hidden conformal symmetry does act locally on the fields in the chiral formulation from \cite{Chen:2025xlo}.
We suspect the same applies to other singleton theories, such as $4d$ linearised gravity, too.

So far, we have concerned ourselves with formulations of free biform singletons.
However, it is interesting to ask whether there exist non-linear theories extending (twisted) self-duality to interactions.
The precedent here is set by spin-1: In this case, while the requirement to respect $\text{U}(1)$ EM duality carves out a specific class of non-linear extensions to free Maxwell theory \cite{Schrodinger:1935oqa,Gaillard:1981rj,bialynicki1983nonlinear,Gibbons:1995cv,Tseytlin:1996it,Gaillard:1997rt,Hatsuda:1999ys,Ivanov:2001ec,Ivanov:2002ab,Ivanov:2003uj,Sorokin:2021tge,Mkrtchyan:2022ulc}, any non-linear theory can be described `democratically' (i.e., involving both dual potentials), even those without duality symmetry \cite{Ivanov:2014nya,Avetisyan:2021heg,Avetisyan:2022zza}.
The situation for spin-2 is more complicated, as gravitational interactions respecting the equivalence principle must be non-Abelian. 
It remains to be seen if there is a non-trivial way around the obstruction (see, e.g., \cite{Monteiro:2023dev}) of extending gravitational duality to interactions.

\section*{Acknowledgments}

The research of CYRC was funded by Grant 115L104044 (Max Planck--IAS--NTU Center) supported by National Taiwan University under the framework of Higher Education SPROUT Project by the Ministry of Education in Taiwan.
The work of EJ is supported by the National Research Foundation of Korea (NRF) grant funded by the Korean government (MSIT) (No. RS-2025-00564305).
KM was supported by UKRI and STFC Consolidated Grants ST/T000791/1 and ST/X000575/1.
KM was also supported by the Higher Education and Science Committee of the Republic of Armenia, under the Remote Laboratory Program, grant number 24RL-1C047.

\appendix

\section{Fibre dualisation and covariant derivatives \label{app: lc coords}}

In this appendix, we collect explicit expressions for fibre dualisation and covariant derivatives in the lightcone basis.

\subsection{Fibre dualisation \label{sapp: fibre dualisation}}

In the lightcone basis, $\{\ul{0},\ul{i}\}$ spans $\mathbb{R}^{1,2p-1}$, acted on by the boundary Lorentz algebra $\mathfrak{so}(1,2p-1)$.

Under dualisation, the components of  an element $B \in \mathcal{V}$ explicitly map as
\begin{subequations}
    \begin{align}
        \spar B_{\ul{-+i_{1} \dots i_{p-1}}} &= \frac{1}{p!}(-)^{p}\, \epsilon_{i_{1} \dots i_{p-1}}{}^{j_{1} \dots j_{p}}\, B_{\ul{0 j_{1} \dots j_{p}}} \\
        \spar B_{\ul{\pm i_{1} \dots i_{p}}} &= \mp \frac{1}{(p-1)!}\,  \epsilon_{i_{1} \dots i_{p}}{}^{j_{1} \dots j_{p-1}}\, B_{\ul{0 \pm j_{1} \dots j_{p-1}}} \\
        \spar B_{\ul{i_{1} \dots i_{p+1}}} &= \frac{1}{(p-2)!} (-)^{p+1}\, \epsilon_{i_{1} \dots i_{p+1}}{}^{j_{1} \dots j_{p-2}}\, B_{\ul{0-+j_{1} \dots j_{p-2}}} \\
        \spar B_{\ul{0 - + i_{1} \dots i_{p-2}}} &= \frac{1}{(p+1)!}\, \epsilon_{i_{1}\dots i_{p-2}}{}^{j_{1} \dots j_{p+1}}\, B_{\ul{j_{1} \dots j_{p+1}}} \\
        \spar B_{\ul{0 \pm i_{1} \dots i_{p-1}}} &= \mp \frac{1}{p!}(-)^{p+1}\, \epsilon_{i_{1} \dots i_{p-1}}{}^{j_{1} \dots j_{p}}\, B_{\ul{\pm j_{1} \dots j_{p}}} \\
        \spar B_{\ul{0 i_{1} \dots i_{p}}} &= \frac{1}{(p-1)!}\, \epsilon_{i_{1} \dots i_{p}}{}^{j_{1} \dots j_{p-1}}\, B_{\ul{+-j_{1} \dots j_{p-1}}},
    \end{align}
    \label{eq: fibre dualisation map}%
\end{subequations}
where the Levi-Civita symbols of $\mathbb{R}^{2,2p}$, $\mathbb{R}^{1,2p-1}$, and $\mathbb{R}^{2p-1}$ are related by
\begin{equation}
    \epsilon_{\ul{-+0i_{1} \dots i_{2p-1}}} = \epsilon_{\ul{0i_{1} \dots i_{2p-1}}} = \epsilon_{\ul{i_{1} \dots i_{2p-1}}},
    \label{eq: epsilon conventions}
\end{equation}
and $\epsilon_{\ul{1 \dots (2p-1)}} = +1$.
We use the same conventions for the spacetime symbols on $\pp M$.
Similar expressions for $B$ can be found by making use of $\spar^{2} = (-)^{p+1}$.
From this we see that $\spar$ maps components with a lower index $\ul{+}$ or $\ul{-}$ to components of the same type --- this fact is used in sections~\ref{sec: cov boundary reduction} and~\ref{sec: non-covariant reduction}.

\subsection{Covariant derivative \label{sapp: covariant derivatives}}

The flatness condition, the connection in lightcone coordinates, and the $J_{\ul{+-}}$-grading~\eqref{eq: bracket -+} are given in the main text; here we present the remaining details.

\paragraph{Generators.}
The generators $J_{\ul{KL}}$ of $\mathfrak{so}(2,2p)$ act on the basis of $\mathbb{R}^{2,2p}$ as
\begin{equation}
    [J_{\ul{KL}}, V_{\ul{I}}] = \eta_{\ul{IL}}\, V_{\ul{K}} - \eta_{\ul{KI}}\, V_{\ul{L}},
\end{equation}
where $\eta_{\ul{IJ}}$ is the flat metric on $\mathbb{R}^{2,2p}$. 
They satisfy
\begin{equation}
    [J_{\ul{KL}},J_{\ul{K'L'}}] = \eta_{\ul{LK'}}\,J_{\ul{KL'}} - \eta_{\ul{LL'}}\,J_{\ul{KK'}} - \eta_{\ul{KK'}}\,J_{\ul{LL'}} + \eta_{\ul{KL'}}\,J_{\ul{LK'}}.
\end{equation}
On the basis of $\mathcal{V}$, they act on each index in turn,
\begin{equation}
    [J_{\ul{KL}}, V_{\ul{I_{1} \dots I_{p+1}}}] = \sum_{n=1}^{p+1} \( \eta_{\ul{I_{n}L}}\, V_{\ul{I_{1} \dots I_{n-1} K I_{n+1} \dots I_{p+1}}} - \eta_{\ul{K I_{n}}}\, V_{\ul{I_{1} \dots I_{n-1} L I_{n+1} \dots I_{p+1}}} \).
    \label{eq: generator commutator}
\end{equation}
Besides the grading~\eqref{eq: bracket -+}, the generators $J_{\ul{\pm\beta}}$ act as
\begin{subequations}
    \begin{align}
        [J_{\ul{\pm\beta}},V_{\ul{\alpha_{1} \dots \alpha_{p+1}}} ] &= (p+1)\eta_{\ul{\beta[\alpha_{1}|}}\,V_{\ul{\pm|\alpha_{2}\dots \alpha_{p+1}]}} \\
        [J_{\ul{\pm\beta}},V_{\ul{\mp \alpha_{1} \dots \alpha_{p}}}] &= -V_{\ul{\beta \alpha_{1} \dots \alpha_{p}}} \pm p \,\eta_{\ul{\beta[\alpha_{1}|}}\,V_{\ul{-+|\alpha_{2} \dots \alpha_{p}]}} \\
        [J_{\ul{\pm\beta}},V_{\ul{\pm \alpha_{1} \dots \alpha_{p}}}] &= 0 \\
        [J_{\ul{\pm\beta}},V_{\ul{+- \alpha_{1} \dots \alpha_{p-1}}}] &= \mp  V_{\ul{\pm \beta \alpha_{1} \dots \alpha_{p-1}}}.
     \end{align}
    \label{eq: bracket +a}
\end{subequations}

\paragraph{Connection.}
For a $\mathcal{V}$-valued form 
\begin{equation}
	X = \frac{1}{(p+1)!}\, X^{\ul{I_{1}\dots I_{p+1}}}\, V_{\ul{I_{1}\dots I_{p+1}}},
\end{equation}
the covariant derivative \eqref{eq: cov div def} reads, in components,
\begin{equation}
    \(\DD X\)_{\ul{I_{1} \dots I_{p+1}}} = \dd X_{\ul{I_{1} \dots I_{p+1}}} - \sum_{n=1}^{p+1}\, \Omega^{\ul{K}}{}_{\ul{I_{n}}} \wedge X_{\ul{I_{1} \dots I_{n-1} K I_{n+1} \dots I_{p+1}}},
\end{equation}
and flatness \eqref{eq: connection flatness} becomes
\begin{equation}
	\dd \Omega^{\ul{I}}{}_{\ul{J}} + \Omega^{\ul{I}}{}_{\ul{K}} \wedge \Omega^{\ul{K}}{}_{\ul{J}} = 0.
\end{equation}
In Poincar\'e coordinates, the non-zero components of $\Omega$ are
\begin{equation}
    \Omega^{\ul{+-}} = -\frac{\dd z}{z},\quad \Omega^{\ul{-\alpha}} = -\sqrt{2}\, \frac{\dd x^{\alpha}}{z},
\end{equation}
in agreement with~\eqref{eq: Omega lightcone}.

\paragraph{Field strength components.} Using this, we find that covariant derivative of $A \in \Omega^{p-1}(M)$ valued in $\mathcal{V}$ has the components
\begin{subequations}
    \begin{align}
        (\DD A)_{\ul{+ \alpha_{1} \dots \alpha_{p}}} &= \dd A_{\ul{+ \alpha_{1} \dots \alpha_{p}}} +  z^{-1}\, \dd z \wedge  A_{\ul{+ \alpha_{1} \dots \alpha_{p}}} \nonumber \\
        &\quad - \sqrt{2}z^{-1}\, \dd x^{\beta}\wedge A_{\ul{\beta \alpha_{1} \dots \alpha_{p}}} - \sqrt{2}p\,z^{-1}\, \dd x_{[\alpha_{1}|} \wedge A_{\ul{-+|\alpha_{2} \dots \alpha_{p}]}} \\
        (\DD A)_{\ul{\alpha_{1}\dots \alpha_{p+1}}} &= \dd A_{\ul{\alpha_{1}\dots \alpha_{p+1}}} +(p+1)\sqrt{2}z^{-1}\, \dd x_{[\alpha_{1}|} \wedge A_{\ul{-|\alpha_{2}\dots \alpha_{p+1}]}} \\
        (\DD A)_{\ul{+- \alpha_{1} \dots \alpha_{p-1}}} &= \dd A_{\ul{+- \alpha_{1} \dots \alpha_{p-1}}} + \sqrt{2} z^{-1}\,  \dd x^{\beta} \wedge A_{\ul{- \beta \alpha_{1} \dots \alpha_{p-1}}} \\
        (\DD A)_{\ul{- \alpha_{1} \dots \alpha_{p}}} &= \dd A_{\ul{- \alpha_{1} \dots \alpha_{p}}}  - z^{-1}\, \dd z \wedge  A_{\ul{- \alpha_{1} \dots \alpha_{p}}}.
    \end{align}
    \label{eq: field strength components}%
\end{subequations}
The components $(\DD A)_{\ul{\alpha_{1} \dots \alpha_{p+1}}}$ and $(\DD A)_{\ul{-+ \alpha_{3} \dots \alpha_{p+1}}}$ are not independent, since they are related by $\spar$.
This is not true for $(\DD A)_{\ul{\pm \alpha_{2} \dots \alpha_{p+1}}}$, which  is mapped to itself under  $\spar$.

\section{Solving the spatial constraints \label{app: constraints}}

In this appendix, we give details on the solution of the constraints in section~\ref{ssec: constraints}.
We write fibre indices first and form indices on $\Sigma$ second, raising and lowering spatial indices with $\delta_{ij}$.
For a field $X$ with $m$ fibre and $n$ form indices, the operations of section~\ref{ssec: constraints} then read
\begin{equation}
    \begin{aligned}
        (\bar{e}\lrcorner X)_{\ul{i_{1}\dots i_{m-1}}\,j_{1}\dots j_{n+1}} &= (n+1)\, X_{\ul{[j_{1}|\, i_{1}\dots i_{m-1}}\,|j_{2}\dots j_{n+1}]},\\
        (\bar{e}\wedge X)_{\ul{i_{1}\dots i_{m+1}}\,j_{1}\dots j_{n+1}} &= (m+1)(n+1)\, \delta_{[\ul{i_{1}}|[j_{1}}\, X_{|\ul{i_{2}\dots i_{m+1}}]\,j_{2}\dots j_{n+1}]}.
    \end{aligned}
    \label{eq: e operations components}
\end{equation}

\subsection{Torsion constraint \label{app: bianchi identity identity}}

With $\phi = 0$, the torsion constraint $\dd Z = \bar{e}\lrcorner\omega$ reads
\begin{equation}
    \pp_{[j_{1}}\, Z_{\ul{i_{1}\dots i_{p-1}}\,j_{2}\dots j_{p}]} = \omega_{\ul{[j_{1}|\, i_{1}\dots i_{p-1}}\,|j_{2}\dots j_{p}]}.
    \label{eq: torsion components}
\end{equation}
Since $Z$ has Young symmetry $[p-1,p-1]$, its transposed derivative $(\dd Z)^{\top}$ has Young symmetry $[p,p-1]$, with the $p$ indices $j$ serving as fibre indices and the $p-1$ indices $i$ as form indices.
We claim that $\omega = (\dd Z)^{\top}$, i.e.\ 
\begin{equation}
	\omega_{\ul{j_{1}\dots j_{p}}\,i_{1}\dots i_{p-1}} = p\,\pp_{[j_{1}}\, Z_{\ul{i_{1}\dots i_{p-1}}\,j_{2}\dots j_{p}]},
\end{equation}
solves \eqref{eq: torsion components}.

Reordering the fibre indices of $\omega$ on the right-hand side of~\eqref{eq: torsion components} and applying \eqref{eq: bianchi identity identity} from the identity below to $Y=\omega$, 
\begin{equation}
	\omega_{\ul{[j_{1}|\, i_{1}\dots i_{p-1}}\,|j_{2}\dots j_{p}]} = \frac{1}{p}\,\omega_{\ul{j_{1}\dots j_{p}}\,i_{1}\dots i_{p-1}},
\end{equation}
where the signs from reordering cancel.
Now, taking $\omega = (\dd Z)^{\top}$, this becomes the left-hand side of \eqref{eq: torsion components}, as required.
For $p=2$, this is the familiar expression of the linearised spin connection in terms of the dreibein.

\begin{identity}
For a tensor $Y$ of Young symmetry $[p,p-1]$, whose first group of $p$ indices we write as fibre indices, we have the identity
\begin{equation}
    Y_{\ul{i_{1} \dots i_{p-1} [i_{p}}\, j_{1} \dots j_{p-1}]} = \frac{1}{p}\,Y_{\ul{j_{1} \dots j_{p-1}i_{p}}\, i_{1} \dots i_{p-1}}.
    \label{eq: bianchi identity identity}
\end{equation}
\end{identity}
\begin{proof}
Note that, by its Young symmetry, any anti-symmetrisation over $p+1$ or more indices on $Y$ vanishes.
This is Bianchi's identity.
We can write down $p-1$ of these:
\begin{subequations}
	\begin{align}
		0 &= (p+1)Y_{\ul{[i_{1} \dots i_{p}} j_{1}] \dots j_{p-1}} \nonumber \\
            &= \binom{p}{0}Y_{\ul{[i_{1} \dots i_{p}}] j_{1} \dots j_{p-1}} +(-)^{p}\binom{p}{1} Y_{\ul{j_{1} [i_{1} \dots i_{p-1}}i_{p} ]j_{2} \dots j_{p-1}}\\
        0 &= (p+1)Y_{\ul{i_{1}[i_{2} \dots i_{p}} j_{1}j_{2}] j_{3}\dots j_{p-1}} \nonumber \\
            &=\binom{p}{1}Y_{\ul{i_{1}[i_{2} \dots i_{p}} j_{1}]j_{2}\dots j_{p-1}}+(-)^{p} \binom{p}{2}Y_{\ul{i_{1}j_{2}[i_{2} \dots i_{p-1}}i_{p} j_{1}]j_{3}\dots j_{p-1}} \\
        &\ \vdots \nonumber \\
        0 &= (p+1)Y_{\ul{i_{1}\dots i_{p-2}[i_{p-1} i_{p}} j_{1}\dots j_{p-1}]} \nonumber \\
            &= \binom{p}{p-2}Y_{\ul{i_{1}\dots i_{p-2}[i_{p-1} i_{p}} j_{1}\dots j_{p-2}]j_{p-1}}+(-)^{p} \binom{p}{p-1}Y_{\ul{i_{1}\dots i_{p-2}j_{p-1}[i_{p-1}} i_{p} j_{1}\dots j_{p-2}]}.
	\end{align}
\end{subequations}
Now, relabelling some of the indices, we then find
\begin{equation}
	\begin{aligned}
		Y_{\ul{i_{1} \dots i_{p-1}[i_{p}}j_{1} \dots j_{p-1}]} &=(-)^{p+1} \frac{p-1}{2}\, Y_{\ul{i_{1} \dots i_{p-2}[i_{p}} j_{1} \dots j_{p-1}]i_{p-1}} \\
		&= (-)^{2(p+1)} \frac{p-1}{2} \frac{p-2}{3}\, Y_{\ul{i_{1} \dots i_{p-3}[i_{p}} j_{1} \dots j_{p-1}]i_{p-2}i_{p-1}} \\
		& \vdots \\
		&= \(\prod_{n=1}^{p-1}(-)^{p+1} \frac{p-n}{n+1}\) Y_{\ul{[i_{p}j_{1} \dots j_{p-1}]}i_{1} \dots i_{p-1}} \\
		&= (-)^{p+1}\frac{1}{p}\, Y_{\ul{[i_{p}j_{1} \dots j_{p-1}]}i_{1} \dots i_{p-1}},
	\end{aligned}
\end{equation}
and moving $i_{p}$ past $j_{1}, \dots, j_{p-1}$ gives a further sign $(-)^{p-1}$, which yields~\eqref{eq: bianchi identity identity}.
\end{proof}

\subsection{Generalised Schouten tensor \label{app: schouten identity}}

The operator $\gamma^{-1}$ of section~\ref{ssec: constraints} acts on fields with $p-1$ fibre and $p-1$ form indices as 
\begin{equation}
	\gamma^{-1} S = *\,\ul{*}\,(\bar{e}\wedge S).
    \label{eq: def gamma inv}
\end{equation}
Using 
\begin{equation}
	\epsilon_{k_{1}\dots k_{p-1}\, i\, i_{1}\dots i_{p-1}}\,\epsilon^{l_{1}\dots l_{p-1}\, i\, j_{1}\dots j_{p-1}} = (2p-2)!\,\delta^{l_{1}\dots l_{p-1} j_{1}\dots j_{p-1}}_{k_{1}\dots k_{p-1} i_{1}\dots i_{p-1}}, 
\end{equation}
with the anti-symmetrised product of Kronecker deltas of unit weight, we find in components
\begin{equation}
    (\gamma^{-1}S)_{\ul{k_{1}\dots k_{p-1}}\,l_{1}\dots l_{p-1}} = \binom{2p-2}{p-1}\, \delta^{l_{1}\dots l_{p-1}\, j_{1}\dots j_{p-1}}_{k_{1}\dots k_{p-1}\, i_{1}\dots i_{p-1}}\; S_{\ul{i_{1}\dots i_{p-1}}\,j_{1}\dots j_{p-1}}.
    \label{eq: gamma inverse}
\end{equation}
This is the generalised Kronecker delta of~\cite{Joung:2012sa}.
Note, in particular, that for $p=2$, $S$ is symmetric and $\gamma^{-1}S = \delta\,\mathrm{tr}\,S - S$, which relates the linearised Schouten and Einstein tensors in three dimensions.

To find the inverse $\gamma$, note that by~\eqref{eq: e operations components}
\begin{equation}
	\bar{e}\wedge S = p^{2}\,\Delta S,
\end{equation}
with
\begin{equation}
    (\Delta S)_{\ul{i_{1}\dots i_{p}}}{}^{j_{1}\dots j_{p}} = \delta_{[\ul{i_{1}}}^{[j_{1}}\, S_{\ul{i_{2}\dots i_{p}}]}{}^{j_{2}\dots j_{p}]}
    \label{eq: schouten identity proxy eq}
\end{equation}
mapping $S$ to a pure-trace tensor with $p$ fibre and $p$ form indices.
Since $*\,\ul{*}$ is an involution, $\gamma = p^{-2}\,\Delta^{-1}\,*\,\ul{*}$, with $\Delta^{-1}$ given in \eqref{eq: schouten identity sol} from the identity below.

For the rotation-curvature constraint $R + 2\,\bar{e}\wedge S = 0$, we have $Y = -R/(2p^{2})$, and this gives the generalised Schouten tensor $S = -\frac{1}{2}\gamma\,G$ in terms of the traces of $R$.

\begin{identity}
The equation 
\begin{equation}
	\delta_{[\ul{i_{1}}}^{[j_{1}}\, S_{\ul{i_{2}\dots i_{p}}]}{}^{j_{2}\dots j_{p}]} = Y_{\ul{i_{1} \dots i_{p}}}{}^{j_{1} \dots j_{p}}
\end{equation}
is solved by
\begin{equation}
    S_{\ul{i_{2} \dots i_{p}}}{}^{j_{2} \dots j_{p}} =  \sum_{n=0}^{p-1} (-)^{n} \[\binom{p}{n+1}\]^{2}\,\delta_{[\ul{i_{2}}}^{[j_{2}} \dots \delta_{\ul{i_{n+1}}}^{j_{n+1}}\, Y^{(n+2)}_{\ul{i_{n+2} \dots i_{p}}]}{}^{j_{n+2} \dots j_{p}]}.
    \label{eq: schouten identity sol}
\end{equation}
\end{identity}
\begin{proof}
For convenience, let us  denote the $[p-n,p-n]$-tensor given by the $(n-1)$-th trace of $S$ by
\begin{equation}
    S^{(n)}_{\ul{i_{n+1} \dots i_{p}}}{}^{j_{n+1} \dots j_{p}} \equiv S_{\ul{k_{2} \dots k_{n} i_{n+1} \dots i_{p}}}{}^{k_{2} \dots k_{n} j_{n+1} \dots j_{p}}.
\end{equation}
In this notation, $S^{(1)}$ is the untraced tensor we are after, and we denote the traces of $Y$ in the same way.

Separating $(i_{n},j_{n})$ from the other indices, we find
\begin{equation}
    \begin{aligned}
        (p-n+1)^{2}\, \delta_{[\ul{i_{n}}}^{[j_{n}}\, S^{(n)}_{\ul{i_{n+1} \dots i_{p}}]}{}^{j_{n+1} \dots j_{p}]}= & \delta_{\ul{i_{n}}}^{j_{n}}\, S^{(n)}_{\ul{i_{n+1}\dots i_{p}}}{}^{j_{n+1} \dots j_{p}} \\
        & \quad + (-)^{p-n}(p-n) \delta_{\ul{i_{n}}}^{[j_{n+1}}\, S^{(n)}_{\ul{i_{n+1} \dots i_{p}}}{}^{j_{n+2} \dots j_{p}]j_{n}} \\
        &\quad + (-)^{p-n}(p-n) \delta_{[\ul{i_{n+1}}}^{j_{n}}\, S^{(n)}_{\ul{i_{n+2} \dots i_{p}}]\ul{i_{n}}}{}^{j_{n+1} \dots j_{p}} \\
        & \quad + (p-n)^{2}\, \delta_{[\ul{i_{n+1}}}^{[j_{n+1}}\, S^{(n)}_{\ul{i_{n+2} \dots i_{p}}]\ul{i_{n}}}{}^{j_{n+2} \dots j_{p}]j_{n}}.
    \end{aligned}
\end{equation}
Taking the trace of this over $(i_{n},j_{n})$, we find
\begin{equation}
    (p-n+1)^{2}\, \delta_{[\ul{i_{n}}}^{[j_{n}}\, S^{(n)}_{\ul{i_{n+1} \dots i_{p}}]}{}^{j_{n+1} \dots j_{p}]}\, \delta^{i_{n}}_{j_{n}} = (2n-1) S^{(n)}_{\ul{i_{n+1} \dots i_{p}}}{}^{j_{n+1} \dots j_{p}} + (p-n)^{2}\, \delta_{[\ul{i_{n+1}}}^{[j_{n+1}}\, S^{(n+1)}_{\ul{i_{n+2}\dots i_{p}}]}{}^{j_{n+2} \dots j_{p}]}.
    \label{eq: trace of antisymmetrised traces}
\end{equation}
Using this,  the $(m+1)$-th trace of the $n=1$ expression is
\begin{equation}
    p^{2}\, \delta_{[\ul{i_{1}}}^{[j_{1}}\, S_{\ul{i_{2} \dots i_{p}}]}{}^{j_{2} \dots j_{p}]}\, \delta^{i_{1}}_{j_{1}} \dots \delta_{i_{m+1}}^{j_{m+1}} = (m+1)^{2}\,S^{(m+1)}_{\ul{i_{m+2} \dots i_{p}}}{}^{j_{m+2} \dots j_{p}} + (p-1-m)^{2}\, \delta_{[\ul{i_{m+2}}}^{[j_{m+2}}\, S^{(m+2)}_{\ul{i_{m+3}\dots i_{p}}]}{}^{j_{m+3} \dots j_{p}]} .
\end{equation}
Since the left-hand side is $p^{2}$ times the corresponding trace of $Y = \Delta S$, we obtain
\begin{equation}
    (m+1)^{2}\, S^{(m+1)}_{\ul{i_{m+2} \dots i_{p}}}{}^{j_{m+2} \dots j_{p}} = p^{2}\, Y^{(m+2)}_{\ul{i_{m+2} \dots i_{p}}}{}^{j_{m+2} \dots j_{p}} - (p-1-m)^{2}\, \delta_{[\ul{i_{m+2}}}^{[j_{m+2}}\, S^{(m+2)}_{\ul{i_{m+3} \dots i_{p}}]}{}^{j_{m+3} \dots j_{p}]},
\end{equation}
which terminates  at $m=p-1$.

Solving iteratively for $S^{(1)}$, we obtain
\begin{equation}
    \begin{aligned}
        S^{(1)}_{\ul{i_{2} \dots i_{p}}}{}^{j_{2} \dots j_{p}} &= p^{2}\, Y^{(2)}_{\ul{i_{2} \dots i_{p}}}{}^{j_{2} \dots j_{p}} - (p-1)^{2}\, \delta_{[\ul{i_{2}}}^{[j_{2}}\, S^{(2)}_{\ul{i_{3} \dots i_{p}}]}{}^{j_{3} \dots j_{p}]} \\
        &= \dots \\
        &=  \sum_{n=0}^{p-1} (-)^{n} \[\binom{p}{n+1}\]^{2}\,\delta_{[\ul{i_{2}}}^{[j_{2}} \dots \delta_{\ul{i_{n+1}}}^{j_{n+1}}\, Y^{(n+2)}_{\ul{i_{n+2} \dots i_{p}}]}{}^{j_{n+2} \dots j_{p}]},
    \end{aligned}
    \label{eq: schouten identity proxy sol}
\end{equation}
as required.
\end{proof}

\bibliographystyle{JHEP}

\bibliography{bibliography.bib}

\end{document}